\documentclass[sn-mathphys-num,iicol]{sn-jnl}

\usepackage{graphicx}%
\usepackage{caption}
\usepackage{subcaption}
\usepackage{lipsum}%
\usepackage{multirow}%
\usepackage{amsmath,amssymb,amsfonts}%
\usepackage{amsthm}%
\usepackage{mathrsfs}%
\usepackage[title]{appendix}%
\usepackage{xcolor}%
\usepackage{cuted}
\usepackage{textcomp}%
\usepackage{manyfoot}%
\usepackage{booktabs}%
\usepackage{algorithm}%
\usepackage{algorithmicx}%
\usepackage{algpseudocode}%
\usepackage{listings}%
\usepackage{makecell}
\usepackage{mathtools}
\usepackage{bbm}
\usepackage{cleveref}
\usepackage{rotating}
\usepackage{mhchem}
\usepackage{xurl}
\usepackage[normalem]{ulem}
\usepackage{siunitx}

\theoremstyle{thmstyleone}%
\theoremstyle{thmstyletwo}%

\theoremstyle{thmstylethree}%

\begin{document}
\title{Tension between MiniBooNE and MicroBooNE within a 3+1 Sterile Neutrino Framework using Simulation-Based Inference}


\author*[1]{\fnm{Julia} \sur{Woodward} \orcid{0009-0006-1636-7562}}\email{julia785@mit.edu}

\author[2]{\fnm{Austin} \sur{Schneider} \orcid{000-0002-0895-3477}}\email{aschneider@tamu.edu}

\author[1]{\fnm{Joshua} \sur{Villarreal} \orcid{0000-0001-9690-1310}}\email{villaj@mit.edu}

\author[1]{\fnm{John} \sur{Hardin} \orcid{0000-0001-8871-8065}}\email{jmhardin@mit.edu}

\author[1]{\fnm{Janet} \sur{Conrad} \orcid{0000-0002-6393-0438}}\email{conrad@mit.edu}

\affil[1]{\orgdiv{Department of Physics}, \orgname{Massachusetts Institute of Technology}, \orgaddress{\street{77 Massachusetts Avenue}, \city{Cambridge}, \postcode{02138}, \state{MA}, \country{USA}}}

\affil[2]{\orgdiv{Department of Physics}, \orgname{Texas A\&M University}, \orgaddress{\street{400 Bizzell St}, \city{College Station}, \postcode{77843}, \state{TX}, \country{USA}}}

\date{\today}% It is always \today, today,

\abstract{The MiniBooNE low-energy excess and its subsequent exclusion by MicroBooNE provide an important test of sterile-neutrino explanations of short-baseline neutrino anomalies. Such a test is complicated by the fact that both experiments use different approaches to their analyses. In this contribution, we present a consistent approach to both experiments. We perform a joint fit of MiniBooNE and MicroBooNE data to a $3+1$ sterile-neutrino model omitting all data-driven factors, using MicroBooNE data from both the Booster Neutrino Beam (BNB) and Neutrinos at the Main Injector (NuMI) beamlines, and evaluate the parameter goodness-of-fit (PG) tension between the two experimental results. A rigorous frequentist treatment of both parameter fitting and PG tension is challenging. Existing methods require asymptotic assumptions known to be inaccurate for neutrino oscillation measurements or repeated likelihood optimization tasks. To this end, we use a previously developed frequentist fitting framework based on simulation-based inference (SBI) and introduce a new SBI-based method for evaluating PG tension that makes the required trial-based calibration computationally feasible. We find that the addition of MicroBooNE reduces the significance of the MiniBooNE preference for sterile-neutrino oscillations, with a trials-based preference for 3+1 remaining at $2.7 \sigma$. The two experiments exhibit a $\geq 2.5\sigma$ PG tension within the $3+1$ model.}


\maketitle

\section{Introduction}\label{sec:introduction}

A number of unexplained anomalies have been observed in neutrino oscillation experiments over the past 25 years. One hypothesis to explain these involves the addition of one or more sterile neutrinos to the Standard Model (SM). Sterile neutrinos do not interact via the weak force but can oscillate to active neutrino flavors, adding additional interference patterns to the expected 3-flavor oscillation probabilities. 

The simplest extension to the SM, called “3+1,” introduces one such sterile neutrino. Evidence for a 3+1 sterile neutrino is measured experimentally via three oscillation channels: $\nu_\mu \rightarrow \nu_e$ (electron-flavor appearance), $\nu_e \rightarrow \nu_e$ (electron-flavor disappearance), and $\nu_\mu \rightarrow \nu_\mu$ (muon-flavor disappearance). It has been well-established that the admission of one additional sterile neutrino to the SM cannot explain all irregularity in the neutrino sector \cite{Hardin:2022muu}, although more complex sterile neutrino models, like 3+2 or 3+3, have historically been too computationally difficult to evaluate. A fit to a 3+1 sterile neutrino model using short-baseline oscillation data finds a $>5 \sigma$ preference for 3+1 over the SM \cite{Hardin:2022muu}.  However, while this quantity measures an overall preference for 3+1, it does not ask if the individual experiments prefer significantly different parameter values, and hence are actually “in tension”. 

The parameter goodness-of-fit (PG) statistic proposed by Maltoni and Schwetz \cite{Maltoni:2003cu} is considered the standard for quantifying tension. The PG tension is defined as the difference between the global and summed likelihoods of $r$ experimental datasets, each with $P_r$ independent parameters: 
\begin{equation}\label{eq:MS_PG}
    \chi^2_{PG} = \chi^2_{glob,min} - \sum \chi^2_{r, min}
\end{equation}
which, under certain requirements similar to Wilks’ theorem, can be assumed to follow a $\chi^2$ distribution with $P_r - P_{glob}$ degrees of freedom. 

It has been known for nearly a decade that there is large internal tension between neutrino oscillation datasets sensitive to anomalous appearance or disappearance.  This tension is driven by differences in the allowed mass splitting between the first three quasi-degenerate mass eigenstates and the fourth mass eigenstate, $\Delta m_{41}^2$. At 95\% confidence, appearance datasets prefer $\Delta m_{41}^2< 1 \text{eV}^2$ while disappearance datasets prefer $\Delta m_{41}^2>6 \text{eV}^2$, resulting in a 4.9$\sigma$ PG tension, as seen in Fig. 11 of Ref. \cite{Hardin:2022muu}. 

Of the experiments sensitive to $\nu_e$ appearance, MiniBooNE exhibits among the highest of 3+1 signal significances, making it a major contributor to the observed tension between appearance and disappearance searches. Located along the Booster Neutrino beam (BNB) line at Fermi National Accelerator Laboratory (Fermilab), MiniBooNE observed a $4.8\sigma$ low energy excess (LEE) of electron-like neutrino events and a 4.6$\sigma$ preference for 3+1 \cite{MiniBooNE:2020pnu, MiniBooNE:2022emn}, consistent with an electron-flavor appearance  fit to data from the LSND experiment reported in 2001 \cite{LSND:2001aii}. 

To investigate the MiniBooNE LEE, the MicroBooNE experiment began operation in 2015 along the same BNB beamline. MicroBooNE also lies off-axis to the Neutrinos at the Main Injector (NuMI) beamline at Fermilab, allowing it to detect neutrinos from both neutrino sources. In 2025, MicroBooNE published a fit to 3+1 using neutrinos produced by both the BNB and NuMI beamlines \cite{MicroBooNE:2025nll}. MicroBooNE reported no evidence for a 3+1 sterile neutrino, in disagreement with MiniBooNE \cite{MicroBooNE:2025nll}. However, the level of disagreement was not quantified by the collaboration. 

Whether the MiniBooNE signal persists at a significant level in a joint analysis with MicroBooNE data from both beamlines is yet unanswered. Furthermore, the degree of underlying tension between the two experiments remains unevaluated, despite its importance for determining whether the 3+1 model can simultaneously describe both datasets, particularly given that the experiments share a beamline.

There are a couple differences between the published 3+1 analysis approaches between the two experiments that must be addressed when answering these questions. For one, the MiniBooNE collaboration drew confidence intervals using Wilks' theorem, which assumes that the test statistic follows a $\chi^2$-distribution with a fixed number of degrees of freedom if certain conditions are met. MicroBooNE instead opted for the frequentist CL$_s$ method, a frequentist procedure specifically designed to set upper limits on new physics models corrected by limitations under systematic uncertainty \cite{Qian:2014nha}. This method is not appropriate to apply in a joint MiniBooNE-MicroBooNE fit where one experiment indicates a signal.

Secondly, the two experiments employ different treatments of data-driven corrections to account for deviations from null and the nominal 3+1 simulation. In the limit of negligible- or zero-value parameters for the proposed additional neutrino state, the 3+1 model reduces to oscillations of the SM's three-neutrino picture (the null hypothesis for later statistical tests). It is possible for data to disagree with both null and 3+1 simulation, in which case analyzers must take care in interpreting the fit results. Such disagreement could also lead to tension observed between experiments even when both prefer null. In this case, both MiniBooNE and MicroBooNE data suffer from excesses compared to prediction in the $\nu_\mu$ channel. The MiniBooNE collaboration therefore introduced a data-driven correction, $f_\pi$, which rescales events originating from charged pion decays to better reproduce the measured charged-current $\nu_\mu$ spectrum \cite{MiniBooNE:2008hfu, Hostert:2024etd}. On the other hand, the MicroBooNE collaboration did not introduce such a factor into their BNB simulation, although MicroBooNE's $\nu_\mu$ BNB data displays an excess at low energies compared to prediction, similar to MiniBooNE's prior to the $f_\pi$ correction. This poses the question of how the tension changes when such post-hoc data-driven factors are included or removed.

A summary of the differences between published 3+1 analyses of MiniBooNE and MicroBooNE is shown in Table \ref{tab:exp_diff}. 

\begin{table*}[ht] 
\centering 
\caption{Comparison of approaches used in collaboration-published 3+1 MiniBooNE and MicroBooNE analyses. Note that all MiniBooNE analyses prior to Ref. \cite{MiniBooNE:2022emn} were appearance-only rather than 3+1.} 
\label{tab:exp_diff} 
\begin{tabular}{lcccc}
\toprule
 \textbf{Dataset (beamline)} & \textbf{Ref.} & \textbf{Significance} & \textbf{$f_\pi$ included} & \textbf{Fitting Method}  \\
\midrule
MiniBooNE (BNB) & \cite{MiniBooNE:2022emn} & $4.6 \sigma$ & \checkmark & Wilks' \\
MicroBooNE (BNB) & \cite{MicroBooNE:2022sdp} & N/A & & Frequentist CL$_s$\\
MicroBooNE (BNB+NuMI) & \cite{MicroBooNE:2025nll} & N/A & & Frequentist CL$_s$ \\
\bottomrule
\end{tabular}
\end{table*}

This paper applies a consistent treatment to both MiniBooNE and MicroBooNE data and is organized as follows. In Sec. \ref{sec:3+1} we introduce the 3+1 model. In Sections \ref{sec:miniboone}-\ref{sec:microboone}, we describe the pair of experiments in detail as well as the pseudo-data generation method to conduct fits. In Sec. \ref{sec:fpiconsistent}, we discuss the relevance of $f_\pi$ in 3+1 fits and in Sec. \ref{sec:wilks-based}, we examine how the inclusion of $f_\pi$ affects both the fit results and the tension between the experiments assuming Wilks' theorem. Sec. \ref{sec:sbi} motivates the use of simulation-based inference (SBI) to permit frequentist fits with fast run-time and introduces an SBI-based analog to the PG tension described earlier. In Sec. \ref{sec:fits}, we present SBI-based fits to the MiniBooNE, MicroBooNE, and combined datasets and evaluate the tension between the two experiments. In Sec. \ref{sec:discussion} we discuss differences between the SBI-based calculations and the Wilks'-based calculations. We conclude in Section \ref{sec:conclusion}.

\section{The 3+1 Model}\label{sec:3+1}
We model each of the three observable oscillation channels for the MiniBooNE and MicroBooNE experiments, in part because an appearance-only approach can lead to an artificially enlarged allowed parameter space \cite{Brdar:2021ysi}. 

If the PMNS matrix is extended with one additional row and column to account for 3+1 oscillations,
\begin{equation}
\begin{pmatrix}
\nu_e\\
\nu_\mu\\
\nu_\tau\\
\nu_s
\end{pmatrix}
=
\begin{pmatrix}
U_{e1} & U_{e2} & U_{e3} & U_{e4}\\
U_{\mu1} & U_{\mu2} & U_{\mu3} & U_{\mu4}\\
U_{\tau1} & U_{\tau2} & U_{\tau3} & U_{\tau4}\\
U_{s1} & U_{s2} & U_{s3} & U_{s4}
\end{pmatrix}
\begin{pmatrix}
\nu_1\\
\nu_2\\
\nu_3\\
\nu_4
\end{pmatrix},
\end{equation}
the oscillation probabilities for each channel are given by:
\begin{eqnarray}
   P_{\nu_\mu \to \nu_\mu} &=& 1-4|U_{\mu 4}|^2 (1- |U_{\mu 4}|^2 ) \sin^2 \Delta_{41} \label{eq:dismu}
\\
    P_{\nu_e \to \nu_e} &=& 1-4|U_{e4}|^2 (1- |U_{e4}|^2 )\sin^2 \Delta_{41} \label{eq:dise}
\\
     P_{\nu_\mu \to \nu_e} &=& 4|U_{\mu 4}|^2|U_{e4}|^2 \sin^2 \Delta_{41} , \label{eq:appe}
\end{eqnarray}
where
\begin{equation}
\Delta_{41}\equiv 1.27 \Delta m^2_{41}L/E. \label{defDelta}
\end{equation}
These probabilities depend on the mass squared splitting between the lightest and heaviest neutrino, $\Delta m^2_{41}=m^2_4-m^2_1$, which sets the frequency of oscillations as a function of neutrino propagation distance, $L$, and neutrino energy, $E$.
The datasets we fit in this work are sensitive to the peak of the first oscillation maximum when $\Delta m^2_{41}\sim 1$ eV$^2$, the case of a ``light" sterile neutrino, large enough to invoke the ``short baseline approximation'' commonly used in global fits, which assumes that the highest mass state is sufficiently heavy so that the three lighter mass states are considered to be degenerate; that is, $\Delta m^2_{41}\approx\Delta m^2_{42}\approx\Delta m^2_{43}$.

The sterile neutrino oscillations in Eqs.~\ref{eq:dismu}, ~\ref{eq:dise}, and ~\ref{eq:appe} can be expressed in terms of angles, and in this work we will make use of the definitions:
\begin{eqnarray}
\sin^2 2\theta_{e e} \equiv 4|U_{e 4}|^2(1- |U_{e4}|^2)
\\
\sin^2 2\theta_{\mu \mu} \equiv 4|U_{\mu 4}|^2(1- |U_{\mu 4}|^2)
\\
\sin^2 2\theta_{\mu e} \equiv 4|U_{\mu 4}|^2|U_{e4}|^2
\end{eqnarray}

Although the mixing matrix may have CP-violating terms, the motivated $\Delta m^2_{41}$ value for a light sterile neutrino leads to negligible sensitivity to CP violation, so the neutrino and antineutrino oscillation probabilities can be taken as identical, even in the case of appearance. 

\section{MiniBooNE}\label{sec:miniboone}
\begin{figure*}[htb]
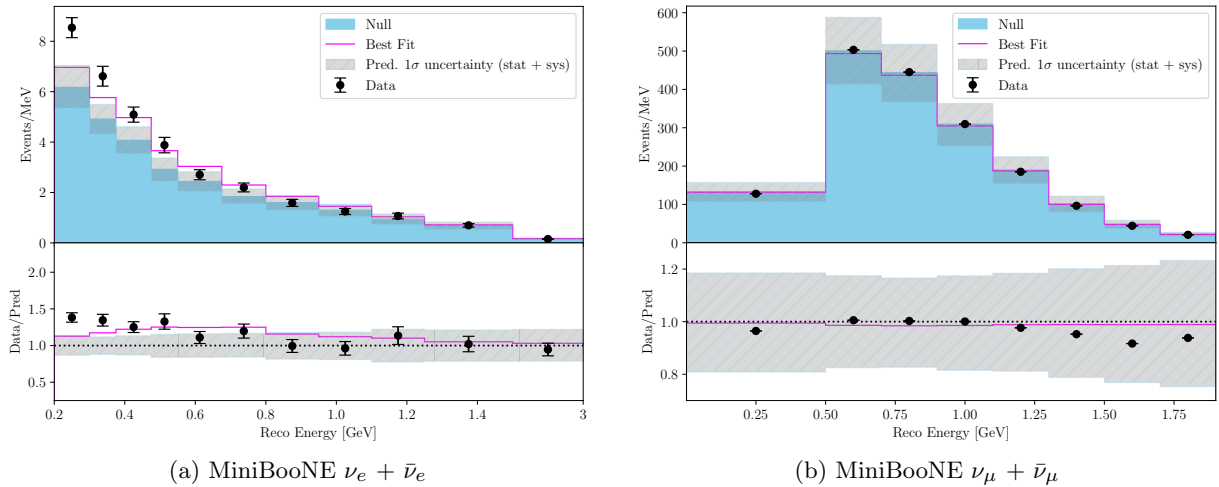

\centering
\begin{subfigure}{0.48\textwidth}
        \centering
        \includegraphics[width=\linewidth]{plots/data/mBnuedata.png}
        
        \caption{MiniBooNE $\nu_e$ + $\bar{\nu}_e$}
        \label{fig:datanuemB}

    \end{subfigure}
    \hfill
    \begin{subfigure}{0.48\textwidth}
        \centering
        \includegraphics[width=\linewidth]{plots/data/mBnumudata.png}
        
        \caption{MiniBooNE $\nu_\mu$ + $\bar{\nu}_\mu$}
        \label{fig:datanumumB}
    \end{subfigure}
    \caption{Data and data/prediction ratio distributions for the MiniBooNE $\nu_e$ and $\nu_\mu$ samples, with the Standard Model prediction shown in blue and the SBI-computed best fit 3+1 prediction in pink. Combined statistical and systematic uncertainty is shown in gray, evaluated from the variance of pseudo-experiments thrown from the published covariance matrix. Statistical error bars are taken from \cite{MiniBooNE:2022emn}.}
    \label{fig:mBdata_dists}
\end{figure*}
MiniBooNE ran from 2002 to 2018 at Fermilab, using an 800t fiducial-volume oil-based Cherenkov detector located 514 m away from the target along
the BNB beamline. The BNB beamline irradiates a Beryllium target with 8 GeV protons to produce primarily pions (with modest kaon content) 
that are collimated by a magnetic focusing horn that can be switched to run in neutrino or antineutrino mode. The decay of these mesons results in a neutrino beam peaked around 800 MeV of mostly muon-flavor, with small electron-flavor content \cite{AGUILARAREVALO200928}.

Prior to 2022, all MiniBooNE analyses were ``appearance-only,'' meaning that only $\nu_\mu \rightarrow \nu_e$ was considered in the analyses and no other Beyond Standard Model effects were assumed \cite{MiniBooNE:2007uho, MiniBooNE:2020pnu}.   
This two-neutrino analysis approach continued through the MiniBooNE collaboration paper of 2021 \cite{MiniBooNE:2020pnu} that presented the final analysis of all collected MiniBooNE data. Inherent in this two-flavor, appearance-only analysis approach is the assumption of negligible  disappearance of either $\nu_\mu$ or $\nu_e$. The $\nu_\mu$ data were solely used to constrain the $\nu_e$ rate, allowing for a data-driven correction to the $\nu_\mu$ flux prediction applied to neutrinos produced by the pion-to-muon decay chain, called $f_\pi$ \cite{MiniBooNE:2008hfu}.  The result is the observed excellent agreement between data and prediction in Fig.~\ref{fig:mBdata_dists}, right.   

Upon applying this factor, and in subsequent analyses,  MiniBooNE reported a consistent $>4 \sigma$ excess of electron-like low energy events compared to prediction, called the ``low energy excess'' (LEE) \cite{MiniBooNE:2018esg}. The excess is shown in Fig. \ref{fig:datanuemB}, left, where the data lie well above the uncertainty band of the prediction at low energies.  The magenta curve on Fig. \ref{fig:datanuemB} shows the best fit to 3+1 oscillations. We observe that the excess is partially, but not completely, explained by the enhancement of electron-flavor appearance by a 3+1 sterile neutrino.   

In 2022, the MiniBooNE collaboration performed a full 3+1 fit, accounting for disappearance effects. The analysis, which also included the $f_\pi$ correction, gave a similar result to the appearance-only fit \cite{MiniBooNE:2022emn}. 
In both cases, with and without disappearance effects, the 3+1 allowed region takes the form of a diagonal band in the $(\sin^2 2\theta_{\mu e}, \Delta m_{41}^2)$ parameter space, corresponding to low $\Delta m^2_{41}$ and high $\sin^2 2\theta_{\mu e}$. Such a preference for low $\Delta m^2_{41}$ solutions is an important driver of the tension in global fits between appearance results and the disappearance results, which prefer high $\Delta m^2$ solutions, and is one of the most important outstanding questions in the field today. 

MiniBooNE also published a 3+1 fit to neutrino events produced by the NuMI beamline, but the data had low statistics and high backgrounds \cite{MiniBooNE:2008hnl}.  Combining fits of the NuMI data set with the MiniBooNE BNB data show marginal change in MiniBooNE's allowed region \cite{Hardin:2022muu}. We therefore do not include this dataset in our analysis. 

To generate pseudo-data for MiniBooNE for fitting in this paper, we make use of the MiniBooNE BNB $\nu_\mu \rightarrow \nu_e$ simulation provided by Ref. \cite{MiniBooNE:2022emn} to perform full 3+1 fits to MiniBooNE data from the BNB beamline.

\section{MicroBooNE}\label{sec:microboone}

\begin{figure*}[htb]
    \centering
    \begin{subfigure}{0.48\textwidth}
        \centering
        \includegraphics[width=\linewidth]{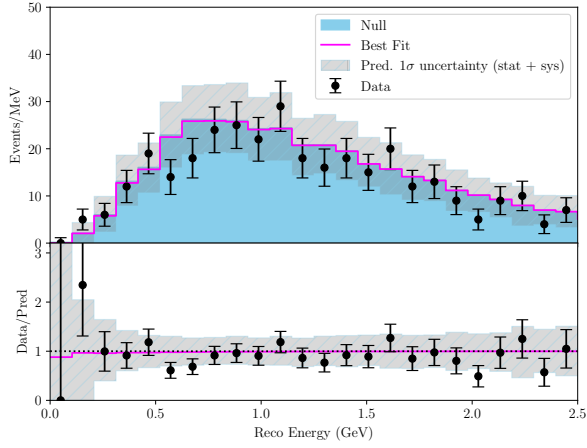}
        
        \caption{MicroBooNE BNB FC $\nu_e$}
        \label{fig:BNBnueFC}
    \end{subfigure}
    \hfill
    \begin{subfigure}{0.48\textwidth}
        \centering
        \includegraphics[width=\linewidth]{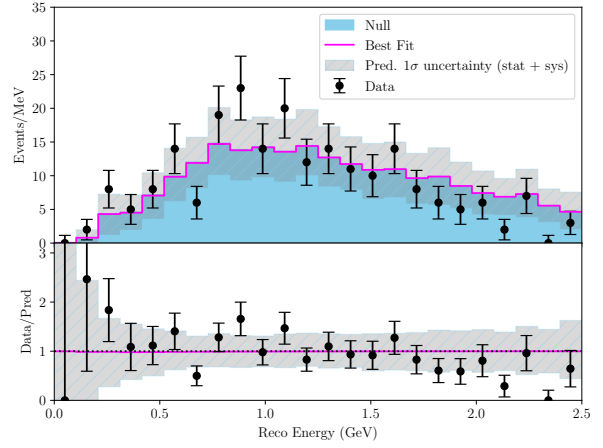}
        
        \caption{MicroBooNE BNB PC $\nu_e$}
        \label{fig:BNBnuePC}
    \end{subfigure}
    \hfill
    \begin{subfigure}{0.48\textwidth}
        \centering
        \includegraphics[width=\linewidth]{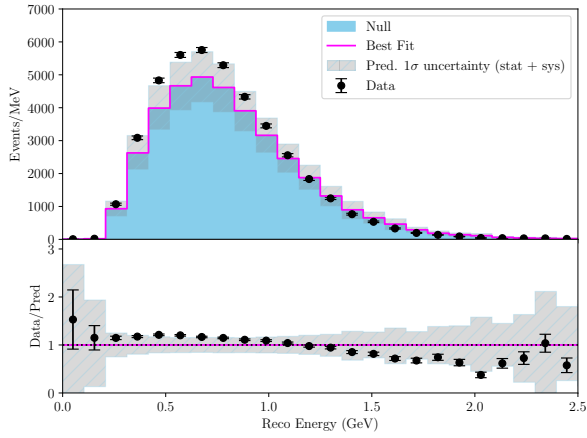}
        
        \caption{MicroBooNE BNB FC $\nu_\mu$ }
        \label{fig:BNBnumuFC}
    \end{subfigure}
    \hfill
    \begin{subfigure}{0.48\textwidth}
        \centering
        \includegraphics[width=\linewidth]{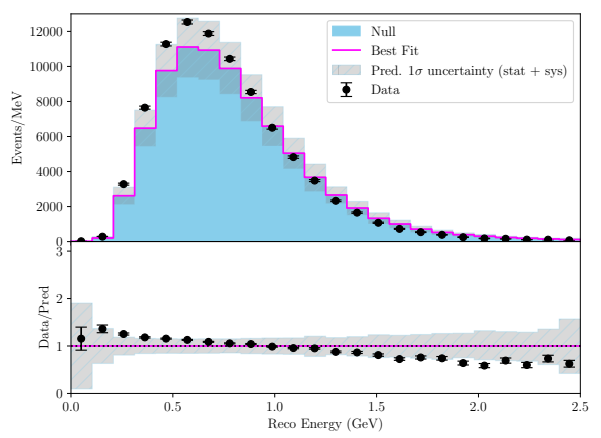}
        
        \caption{MicroBooNE BNB PC $\nu_\mu$ }
        \label{fig:BNBnumuPC}
    \end{subfigure}
    \hfill
    \caption{Data and data/prediction ratio distributions for all MicroBooNE BNB samples used in the analysis, with the Standard Model prediction shown in blue and the SBI-computed best fit 3+1 prediction in pink. Combined statistical and systematic uncertainty is shown in gray, evaluated from the variance of pseudo-experiments thrown from the published covariance matrix. Statistical error bars are taken from \cite{MicroBooNE:2025nll}.}
    \label{fig:BNBdata_dists}
\end{figure*}

The MicroBooNE detector sits upstream of MiniBooNE, 470m away from the production target of the BNB beamline, as well as 8$^\circ$ off-axis and 679m away from the NuMI beamline. MicroBooNE collected data
from 2015 to 2020 with the primary physics goal of investigating the LEE. Its detector consists of a 170-tonne Liquid Argon Time Projection Chamber, giving it excellent
reconstruction capabilities to reduce particle misidentification to which MiniBooNE was susceptible. The difference in the shape and magnitude of the predicted energy spectra between MicroBooNE and MiniBooNE seen in Figs. \ref{fig:datanuemB} and \ref{fig:BNBnueFC}/\ref{fig:BNBnuePC}  reflect the differences in reconstruction capabilities. The
two experiments are thus complementary to one another.  Results of a 3+1 joint analysis of only BNB beamline data have already been published in Ref. \cite{MiniBooNE:2022emn}.

In 2022, MicroBooNE published a 3+1 fit to BNB-only data, finding consistency with the Standard Model \cite{MicroBooNE:2022sdp}. The fit yielded a  $\chi^2/\mathrm{dof}$ of $ 86.62/179$, corresponding to a $p$-value of approximately $1-5\times 10^{-10}$. Such a low $\chi^2$ indicates that the observed deviations between data and prediction are, on average, substantially smaller than expected under the assumed covariance matrix. It is impossible for analyzers external to the collaboration to comment on the source of this behavior. In Appendix \ref{sec:BNBfpi}, we reproduce this behavior and encourage caution when interpreting a goodness-of-fit this far below unity.

In 2025, MicroBooNE published results using data from both BNB and NuMI beamlines, and also found consistency with the Standard Model \cite{MicroBooNE:2025nll}. The NuMI beam produces neutrinos by colliding 120 GeV protons -- substantially more energetic than the protons used for the BNB beam -- at a graphite target. Two pulsed magnetic ``horns'' focus the produced pions and kaons, whose decays make the neutrinos. Unlike the case of MiniBooNE, the NuMI data set is statistically strong enough to break some of the degeneracy between $\nu_e$ appearance and $\nu_e$ disappearance in the BNB data set.  This is due to different ratios of intrinsic $\nu_e$ to $\nu_\mu$ content in each beam. Therefore, we include both data sets in our analysis.  
 
Figures \ref{fig:BNBdata_dists} and \ref{fig:NuMIdatadists} present energy spectra for all MicroBooNE BNB and NuMI samples used in our analysis, consisting of $\nu_e$ and $\nu_\mu$ fully contained (FC) and partially contained (PC) samples. We note that the SM predictions shown are from first-principles, meaning they are not constrained by any fit to the 3+1 model. We omit the $\pi^0$ samples from our analysis because sufficient energy reconstruction information was not released for those samples, and because Ref. \cite{MiniBooNE:2022emn} omitted the same constraints for the MicroBooNE BNB-only fit and achieved strong agreement with published collaboration results. Our NuMI prediction is generated with fixes later discussed in Sec \ref{sec:NuMIsim}, thus, there may be small deviations from published spectra. 

There are two features in the data that affect a 3+1 analysis. First, in the BNB $\nu_e$ FC channel, there is an overall systematic data-deficit above 500 MeV, although it lies within the predicted $1 \sigma$ uncertainty bands. Second, in the $\nu_\mu$ channel for both beamlines, there is a notable excess of events at lower energies and a downward slope in the data/prediction ratio, particularly visible for the BNB data.  We discuss the implications of the excesses in Sec.~\ref{sec:fpisource}.

This work makes use of the MicroBooNE 2025 data release \cite{MicroBooNE:2025nll}.   We do not use Bin 26 of the data release as its width is not clearly defined, as confirmed by MicroBooNE collaborators \cite{uBprivate}. Because the release does not provide event-level Monte Carlo information, the BNB and NuMI predictions cannot be reproduced directly. We provide details on the generation of MicroBooNE BNB and NuMI predictions below.
 
\subsection{BNB prediction}

A MicroBooNE $\nu_\mu \rightarrow \nu_e$ simulation is not currently available from public data releases. Instead, the MicroBooNE $\nu_e$ prediction is obtained using the MiniBooNE BNB simulation with modified baseline to compute a ratio between the nominal $\nu_e$ intrinsic background prediction and the $\nu_e$ prediction with oscillation effects as a function of true neutrino energy~\cite{MiniBooNE:2020pnu_HEPData}. The intrinsic $\nu_e$ prediction obtained from released MicroBooNE $\nu_e$ simulation is multiplied by this ratio and then added to a constant background to obtain the total prediction in the electron flavor channel. The result of this procedure was shown to agree well with the MicroBooNE publications (Fig. S2 in Ref. \cite{MiniBooNE:2022emn}).

\subsection{NuMI prediction \label{sec:NuMIsim}}

\begin{figure*}[htb]
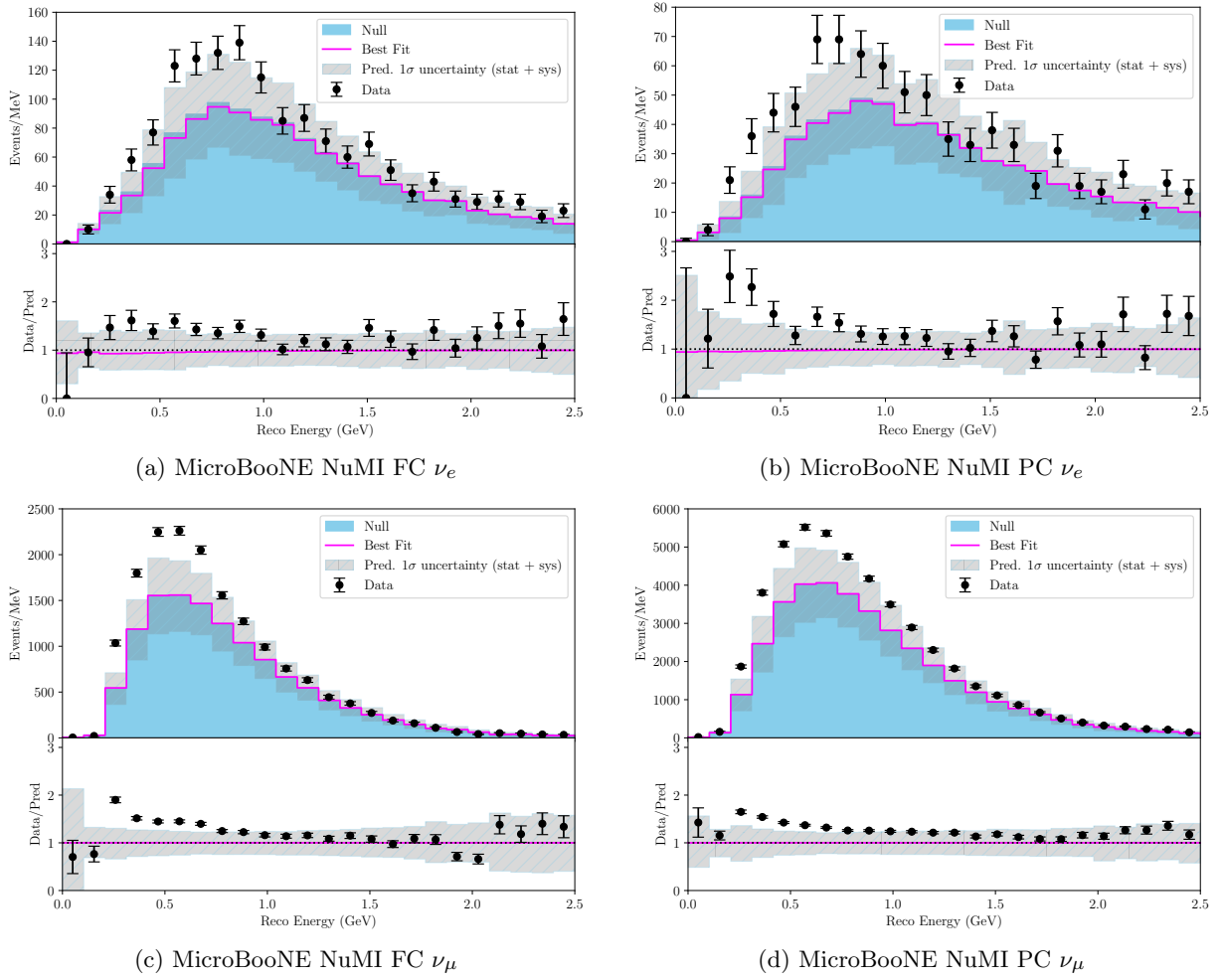

\centering
\begin{subfigure}{0.48\textwidth}
        \centering
        \includegraphics[width=\linewidth]{plots/data/NuMInue_FC_data.png}
        
        \caption{MicroBooNE NuMI FC $\nu_e$}
        \label{fig:NuMInueFC}
    \end{subfigure}
    \hfill
    \begin{subfigure}{0.48\textwidth}
        \centering
        \includegraphics[width=\linewidth]{plots/data/NuMInue_PC_data.png}
        
        \caption{MicroBooNE NuMI PC $\nu_e$}
        \label{fig:NuMInuePC}
    \end{subfigure}
    \hfill
    \begin{subfigure}{0.48\textwidth}
        \centering
        \includegraphics[width=\linewidth]{plots/data/NuMInumu_FC_data.png}
        
        \caption{MicroBooNE NuMI FC $\nu_\mu$}
        \label{fig:NuMInumuFC}
    \end{subfigure}
    \hfill
    \begin{subfigure}{0.48\textwidth}
        \centering
        \includegraphics[width=\linewidth]{plots/data/NuMInumu_PC_data.png}
        
        \caption{MicroBooNE NuMI PC $\nu_\mu$}
        \label{fig:NuMInumuPC}
    \end{subfigure}
    \caption{Same as Fig. \ref{fig:BNBdata_dists} for NuMI data samples.}
    \label{fig:NuMIdatadists}
\end{figure*}

Although the data release associated with MicroBooNE's BNB+NuMI sterile neutrino analysis~\cite{MicroBooNE:2025nll} includes other information needed for the fit, it does not contain per-simulation-event neutrino energies and propagation distances needed for weighting to different neutrino oscillation scenarios.
At the time of writing, this information is not available in any public MicroBooNE data release.
To obtain the per event information necessary for the oscillation predictions we use the same procedure as MicroBooNE:
\texttt{g4numi} simulates proton interactions up until neutrino production~\cite{g4numi_software,Adamson:2015dkw,Pavlovic:2008zz,Loiacono:2010zza,AliagaSoplin:2016shs}, \texttt{PPFX} reweights interactions~\cite{MINERvA:2016iqn} from the \texttt{g4numi} stage, \texttt{dk2nu} biases neutrino production towards the detector~\cite{Hatcher:2012dk2nu,hatcher2025dk2nu}, and neutrino interactions are modeled with \texttt{GENIE}~\cite{Andreopoulos:2009rq,Andreopoulos:2015wxa,GENIE:2021zuu,GENIE:2021wox}.
In lieu of a detector simulation we tune an approximate detector response model to match MicroBooNE predictions, efficiencies, and reconstruction smearing.

\texttt{g4numi} is a Geant4 simulation that begins from the initial \qty{120}{GeV} proton, models proton interactions on the graphite target, and all subsequent interactions until neutrino production through meson- or muon-decay~\cite{GEANT4:2002zbu,Allison:2006ve,Allison:2016lfl}.
We are able to reproduce MicroBooNE's neutrino flux predictions~\cite{Mistry:2021xsb,MicroBooNE:2024numiflux} to within the reported Monte Carlo statistical uncertainties using the same \texttt{g4numi}, Geant4, \texttt{PPFX}, and \texttt{dk2nu} versions~\cite{nayak2024g4numi_uboone,Geant4:v10.4.2,nayak2024ppfx,hatcher2025dk2nu,nayak2026numiflux}.
This reproduced flux simulation is used to tune our detector response model, while we use a modified version of \texttt{g4numi} for the rest of this work that has several improvements not included in the MicroBooNE BNB+NuMI sterile neutrino analysis~\cite{MicroBooNE:2025nll}.
Our modified \texttt{g4numi} simulation~\cite{schneider2026g4numi} fixes a backwards momentum cut that was removing more than half of the decay-at-rest neutrino flux; for reconstructed energies below \qty{250}{MeV} this issue reduced the rate by up to \qty{10}{\%} and has been reported to and confirmed by the MicroBooNE collaboration.
We also include a fix to overlapping geometry elements of the focusing horn and the surrounding shielding.
This issue, identified and fixed by MicroBooNE collaborators, resulted in a $3-4\%$ over-prediction of the muon-neutrino flux from \qty{250}{MeV}-\qty{3}{GeV} for FHC and between \qty{1}{GeV}-\qty{3}{GeV} for RHC.
The resulting fluxes per cm$^2$ per POT are given in Table~\ref{tab:microboone_flux}.

To predict the event rates, we must also include the neutrino interaction cross section on $^{40}$Ar, the detection efficiency, and the smearing from true energy to reconstructed energy.
We use the untuned \texttt{G18\_10a\_02\_11a} cross section from \texttt{GENIE} v3.0.4~\cite{genie_xsec_v3_00_04_ub2}; this differs from MicroBooNE's tuned GENIE cross sections~\cite{MicroBooNE:2021ccs} but the difference is absorbed by an energy-dependent efficiency factor.
Event migration and reconstruction smearing is modeled by column-normalized versions of smearing matrices from the data release of MicroBooNE's Wire-Cell BNB analysis~\cite{MicroBooNE:2021nxr_HEPData,MicroBooNE:2021nxr} split by channel ($\nu_\mu$-FC/$\nu_\mu$-PC/$\nu_e$-FC/$\nu_e$-PC); this assumes that reconstruction smearing is similar between the BNB and NuMI samples, and can be sufficiently described only as a function of the sample and true neutrino energy.
Detection efficiency is modeled separately for each of the four samples as a function of true neutrino energy by ten-knot linear interpolation, introducing 10 free parameters per sample that must be constrained.
These parameters are optimized by fitting the reconstructed energy spectrum to the official MicroBooNE predictions for the NuMI reconstructed energy spectrum~\cite{MicroBooNE:2025nll} with a prior to maintain smooth curvature.
The $\nu_e$ samples have additional priors driving them towards the shape-released efficiency curves of the BNB sample~\cite{MicroBooNE:2021nxr_HEPData} with the normalization left free and a prior to maintain monotonicity, while the $\nu_\mu$ samples include priors to prefer a single-peak efficiency curve with both a low-energy threshold shape and mono-tonic high-energy tail.
The BNB data release constrains the detector response only up to \qty{3}{GeV}, whereas the NuMI flux extends well beyond it.
Above \qty{3}{GeV} we extrapolate: the efficiency is set to its \qty{3}{GeV} value but scaled by a containment fall-off from a geometric model that propagates \texttt{GENIE}-generated charged-current kinematics through the TPC, and the energy migration is taken from the \qty{3}{GeV} column of the published response matrices.
Below about \qty{250}{MeV}, where the BNB response matrix has poor statistical support, we use a phenomenologically-motivated smearing model that is fit simultaneously to the columns of the BNB response matrices and to the NuMI predictions.

\begin{table*}[tb]
\centering
\caption{Our predicted fluxes at MicroBooNE above 60~MeV, assuming no oscillation, in neutrinos per cm$^2$ per POT.}
\label{tab:microboone_flux}
\begin{tabular}{lcc}
\hline
Species & FHC flux $[\unit{cm^{-2}}~\unit{POT^{-1}}]$ (fraction) & RHC flux $[\unit{cm^{-2}}~\unit{POT^{-1}}]$ (fraction) \\
\hline
$\nu_\mu$       & $5.4\times10^{-10}$ (66\%)  & $3.2\times10^{-10}$ (42\%) \\
$\bar{\nu}_\mu$ & $2.6\times10^{-10}$ (32\%)  & $4.3\times10^{-10}$ (56\%) \\
$\nu_e$         & $1.3\times10^{-11}$ (1.6\%) & $8.9\times10^{-12}$ (1.2\%) \\
\hline
\end{tabular}
\end{table*}

\section{Consistent treatment of data-driven corrections ($f_\pi$)}\label{sec:fpiconsistent}

As discussed in Secs.~\ref{sec:miniboone} and \ref{sec:microboone}, both MiniBooNE and MicroBooNE have observed notable $\nu_\mu$ excesses within their respective systematic assumptions. The two analyses, however, treat the underlying flux normalization differently: MiniBooNE applies a data-driven correction, $f_\pi$, whereas MicroBooNE reports its $3+1$ analysis without an analogous correction despite sharing the same beamline. A joint fit should require a consistent treatment of $f_\pi$. Although global fits ordinarily use the data and systematic treatment presented by each collaboration \cite{Hardin:2022muu, Diaz:2019fwt}, this case warrants additional consideration because MiniBooNE attributed the underestimated prediction to the normalization of pion production in the beam \cite{Schmitz:2008zz}. If this interpretation is correct, the corresponding systematic effect should be common to both experiments, motivating a consistent treatment of MiniBooNE and MicroBooNE in the joint analysis \cite{Hostert:2024etd}.   

The simplest and most model-agnostic approach is to remove the $f_\pi$ correction from the MiniBooNE prediction. In this work, all MiniBooNE fits presented will have $f_\pi$ removed unless otherwise noted. 

\subsection{Understanding the source of the $f_\pi$ correction across MiniBooNE and MicroBooNE datasets}\label{sec:fpisource}

As noted above, the best explanation for the $\nu_\mu$ underprediction in MiniBooNE appeared to be mismodeling of the rate of pion production in the BNB beam \cite{Schmitz:2008zz}. It should be noted that MiniBooNE neutrino data and antineutrino data gave nearly identical results for this corrective factor (Table \ref{tab:fpi}). This neutrino-antineutrino $f_\pi$ agreement could point to a normalization issue at the proton interaction point. 

\begin{table}[ht] 
\centering 
\caption{Factors of $f_\pi$ applied to events coming from charged pion decay in the MiniBooNE MC.} 
\label{tab:fpi} 
\begin{tabular}{lcc}
\toprule
 & \textbf{Neutrino Mode} & \textbf{Antineutrino Mode} \\
\midrule
$\pi^+$ & 1.28 & 1.00 \\
$\pi^-$ & 0.98 & 1.28 \\
\bottomrule
\end{tabular}
\end{table}

However, the MicroBooNE data (Ref.  \cite{MicroBooNE:2025nll}) raises questions about potential proton-interaction-based explanations. Adding an $f_\pi$ factor to MicroBooNE does not completely match data with prediction (see Appendix \ref{sec:BNBfpi}), because there is an energy dependence to the MicroBooNE $\nu_\mu$ excess, as depicted in the data/MC ratio distributions from Figs. \ref{fig:BNBnumuFC} and \ref{fig:BNBnumuPC}.  Such an energy dependence is not observed in MiniBooNE data. The difference in energy dependence between MiniBooNE and MicroBooNE BNB results could point to an issue with nuclear effects or cross sections that lead to slightly different normalizations or energy dependencies between argon (MicroBooNE) and carbon (MiniBooNE).

We also note that the MicroBooNE NuMI data presents a considerable $\nu_\mu$ excess (Fig. \ref{fig:NuMIdatadists}, bottom). However, this discrepancy cannot be straightforwardly attributed to pion production mismodeling, since the NuMI beam has a much higher kaon content compared to the BNB beam, indicating the effect is likely not solely from pions.

We conclude that the source of the underprediction is not well-understood, and should not be assumed to be in the beamline.  However, because the $\nu_\mu$ prediction is highly correlated to the $\nu_e$ background prediction by the pion decay chain, the data-MC $\nu_\mu$ normalization disagreement needs to be identified and modeled. Until this mis-estimation is identified, quantitative conclusions comparing the experiments should be considered with care.

\subsection{Methods for removing $f_\pi$}

To remove $f_\pi$ from the MiniBooNE fit, we make use of parent tagging in the MiniBooNE MC to identify the $\nu_e/\bar{\nu}_e$ and $\nu_\mu/\bar{\nu}_\mu$ events originating from a charged pion decay and undo the applied $f_\pi$ scaling for those events (see Table \ref{tab:fpi}).

\begin{figure*}[htb]
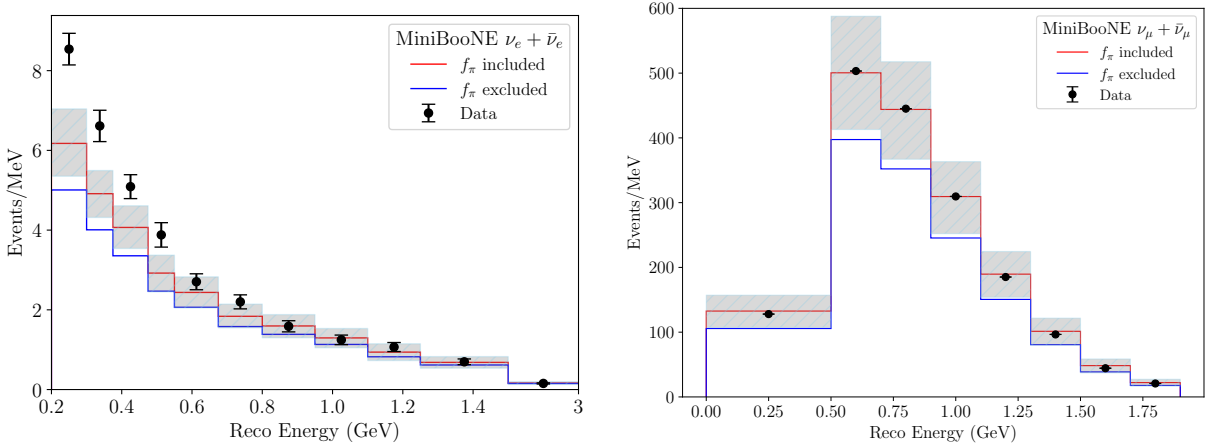

    \centering
    \begin{subfigure}{0.48\textwidth}
        \includegraphics[width=\linewidth]{plots/data/mBnuefpi_compare.png}
    \end{subfigure}
    \hfill
    \begin{subfigure}{0.48\textwidth}
        \includegraphics[width=\linewidth]{plots/data/mBnumufpi_compare.png}
    \end{subfigure}
    \hfill
    \caption{Null MiniBooNE expectation for $\nu_e$ and $\nu_\mu$ samples with and without $f_\pi$ included in the fit. Predicted $1 \sigma$ combined statistical and systematic uncertainty bands for the prediction with $f_\pi$ included (see Fig. \ref{fig:mBdata_dists}) are shown in gray.}
    \label{fig:fpidists}
\end{figure*}

Reconstructed energy distributions at null with and without the factors of $f_\pi$ applied are shown in Figure \ref{fig:fpidists}. The inclusion of $f_\pi$ pulls up the prediction in both the $\nu_e$ and $\nu_\mu$ channel towards the data, improving the fit to null, but not eliminating the LEE entirely. The $\nu_\mu$ data is now in almost perfect agreement with prediction, and the low energy excess in the $\nu_e$ channel decreases significantly. 

\section{Impact of $f_\pi$ on Wilks'-based fits and tensions}\label{sec:wilks-based}

In this subsection, we investigate how the MiniBooNE and joint fits change with the inclusion of $f_\pi$ in MiniBooNE. In Appendix \ref{sec:BNBfpi}, we investigate how the fits change when $f_\pi$ is included in MicroBooNE BNB-only predictions.

Acknowledging that both experiments opted for different fitting methods, to enable a consistent comparison of signal-like and exclusion-like data, we apply the same Wilks’-based procedure to both MiniBooNE and MicroBooNE in this section. In Sec. \ref{sec:sbi}, we instead propose a consistent frequentist fitting method between the two experiments using simulation-based inference. This allows cross-comparison of the fit results between the two experiments using different fitting approaches.

\begin{figure}
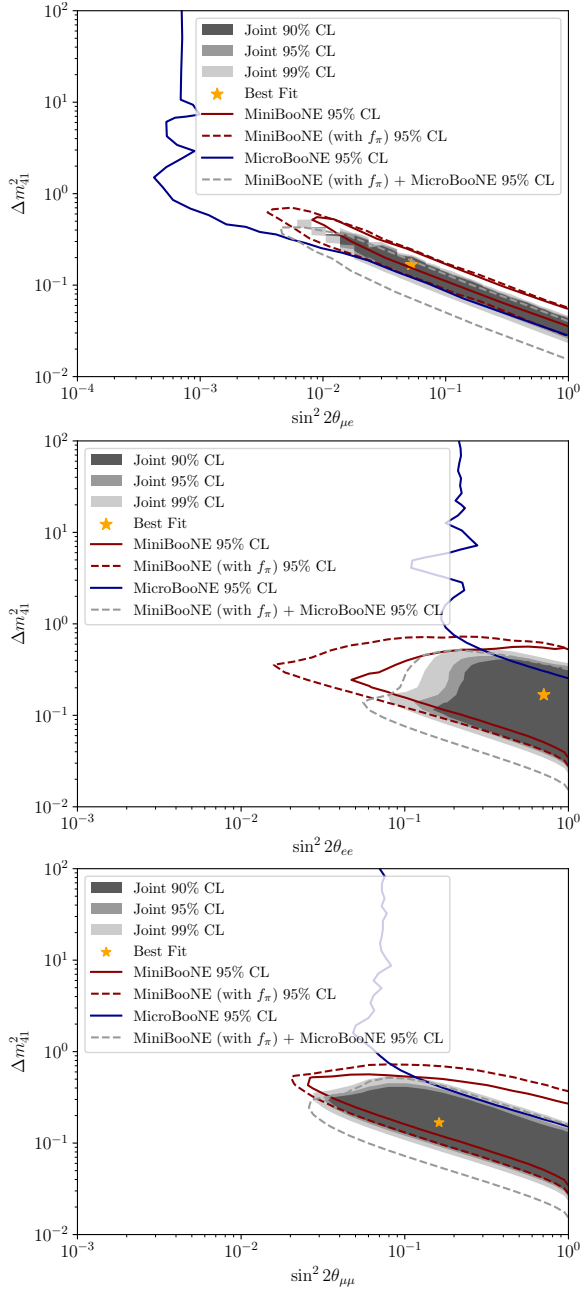

    \centering
    \begin{subfigure}{\linewidth}
        \includegraphics[width=\linewidth]{plots/fits/Wilks_sin22th_real_data_comb.png}
    \end{subfigure}
    \hfill
    \begin{subfigure}{\linewidth}
        \includegraphics[width=\linewidth]{plots/fits/Wilks_sin22thee_real_data_comb.png}
    \end{subfigure}
    \hfill
    \begin{subfigure}{\linewidth}
        \includegraphics[width=\linewidth]{plots/fits/Wilks_sin22thmumu_real_data_comb.png}
    \end{subfigure}
    \hfill
    
    \caption{Wilks'-based confidence regions from a joint fit to MiniBooNE (without $f_\pi$ corrections) and MicroBooNE data. Each fit shown profiles over the third parameter to obtain the likelihood. Overlaid are the MicroBooNE-only 95\% CL (blue, solid), MiniBooNE-only 95\% CL (red, solid), the MiniBooNE-only 95\% CL including $f_\pi$ (red, dashed),  and the joint MiniBooNE (including $f_\pi$) and MicroBooNE 95\% CL (grey, dashed).}
    \label{fig:wilks-combfits}
\end{figure}

In Figure \ref{fig:wilks-combfits}, we present a joint fit to MiniBooNE (without $f_\pi$) and MicroBooNE (BNB+NuMI) data assuming Wilks' theorem for each of the three relevant oscillation channels, with single-experiment comparisons overlaid. Traditionally, 3+1 results are shown in two-dimensional plots after profiling over the third parameter, as is presented here. However, profiling averages over information that may be important in the complete global fit \cite{Rodrigues:2025tha}, therefore the three-dimensional results are presented in Appendix \ref{app:wilks3D}.

For the single-experiment fits that adopt the $f_\pi$ treatment reported by the respective collaborations---MiniBooNE with $f_\pi$ and MicroBooNE without---we find good agreement with published results \cite{MiniBooNE:2022emn, MicroBooNE:2025nll}. This represents an important cross check for the joint fit that is presented for the first time in this work. This also means the 90\% exclusion for MicroBooNE is only marginally different from the Wilks’ method as opposed to CL$_s$. 

From Fig. \ref{fig:wilks-combfits}, the Wilks'-based joint fit (without $f_\pi$ corrections in MiniBooNE) is topologically similar to the MiniBooNE-only fit, while shifted toward slightly lower values in $\Delta m_{41}^2$, as is evident in the $\nu_e$ appearance plot. This behavior is expected, since the MicroBooNE data shows no evidence of $\nu_e$ excess at 2$\sigma$ over most of the MiniBooNE 95\% CL allowed region. We note that when $f_\pi$ is included in the MiniBooNE fit, the joint fit allowed region is shifted even farther downward from the MiniBooNE-only fit since the MiniBooNE signal is not as strong and thus does not dominate the fit as heavily. Similarly, including $f_\pi$ widens the MiniBooNE-only confidence regions, corresponding to a reduction in MiniBooNE's signal significance.

\begin{table*}
    \centering
    \caption{Statistical significance and tension quoted in $\sigma$ for each dataset computed using an analytic $\chi^2$ assuming a $\chi^2_{dof=3}$ distribution provided by Wilks' theorem.  ``$f_\pi$ Included/Excluded" refers to the inclusion of $f_\pi$ corrections in the MiniBooNE fit.}
    \label{tab:significances_wilks}
    
    \begin{tabular*}{\linewidth}{@{\extracolsep{\fill}} lcc}
        \toprule
        & \multicolumn{2}{c}{\textbf{Significance in $\sigma$}} \\
        \cmidrule(lr){2-3}
        \textbf{Dataset(s)} & $f_\pi$ Included & $f_\pi$ Excluded \\
        \midrule
        MiniBooNE &  4.8 &  7.1\\
        MiniBooNE + MicroBooNE & 3.0 &  5.2\\
        
        \midrule
        & \multicolumn{2}{c}{\textbf{Tension in $\sigma$}} \\
        \cmidrule(lr){2-3}
        \textbf{Dataset(s)} & $f_\pi$ Included & $f_\pi$ Excluded \\
        \midrule
        MiniBooNE v. MicroBooNE & 3.2 & 4.4\\
        
        \bottomrule
    \end{tabular*}
    \vspace{0.5em}

\end{table*}

The signal significances and tension between the MiniBooNE and MicroBooNE datasets are shown in Table \ref{tab:significances_wilks}. As expected from Fig. \ref{fig:fpidists}, the difference in MiniBooNE’s signal significance with and without $f_\pi$
corrections indicates much stronger disagreement with null when $f_\pi$ is excluded. A $7 \sigma$ signal is enormous and often implies the data are outside the valid range of uncertainty estimation. We also caution that such a strong oscillatory signal does not fall into the Wilks' regime, and so the significance must be interpreted with care \cite{algeri2020searching, Hardin:2022qdh}. In both the MiniBooNE-only fit and the joint fit, the removal of $f_\pi$ raises the signal preference for 3+1 by about $2 \sigma$, and raises the tension between the two experiments by about $1 \sigma$. This is because a combined fit struggles to find a consistent set of parameters that fits both datasets sufficiently, due to the increased MiniBooNE $\nu_e$ excess and lack of MicroBooNE $\nu_e$ excess. 

We also note that in both $f_\pi$ cases (included or omitted), the addition of MicroBooNE data from both beamlines lowers the signal significance substantially by about $2 \sigma$, and the tensions indicate statistically significant disagreement between the two datasets.

\section{Simulation Based Inference}\label{sec:sbi}
In this section, we motivate the use of simulation-based-inference in global fits and describe its utility in both frequentist fitting and tension evaluation. 

In maximum likelihood inference, Wilks' theorem is often assumed. Unfortunately, when used to fit sinusoidal signals like neutrino oscillations, qualifying criteria for Wilks' theorem do not hold \cite{algeri2020searching}, causing the construction of confidence levels using this method to fail to satisfy expected coverage properties, impacting fit interpretability \cite{Hardin:2022qdh}. A domain-standard solution is to use the trials-based Feldman-Cousins method \cite{feldman-cousins}, which evaluates a large number of pseudoexperiments generated from each parameter space point to calculate critical cutoff values empirically. This is often computationally infeasible due to the repeated likelihood optimization tasks required. 

Instead, simulation based inference (SBI) approaches of CL construction have proven to alleviate computational costs while maintaining expected coverage properties. SBI is a subfield of machine learning in which the goal is to learn statistical properties of simulated data to enable rapid inference on observations \cite{doi:10.1073/pnas.1912789117}. Such approaches are particularly valuable when the likelihood function is intractable or too complex to optimize efficiently, as is often the case in global fits of sterile neutrino oscillations. In previous work, we have demonstrated that SBI can be used for trials-based global fits, resulting in a four order of magnitude faster evaluation per grid point than the method of Feldman and Cousins \cite{10.1088/2632-2153/ae040c}. The framework for performing such a fit is outlined in Ref. \cite{10.1088/2632-2153/ae040c}. To benchmark the performance of this approach, we demonstrate the calibration of SBI-based fitting and tension evaluations in Appendix \ref{sec:calibration} on a toy experiment sensitive to electron neutrino appearance.

\subsection{Likelihood Ratio Estimation with Simulation-Based Inference}
In the work presented here, we adopt methods presented in Ref.~\cite{10.1088/2632-2153/ae040c} to perform a frequentist fit to data using SBI. While described in detail therein, we summarize those techniques here for completeness.

To this end, an analyzer trains a neural network classifier on simulated data to infer the likelihood ratio for pseudo-experimental realization $x$ between two physics parameters $\theta_1 \equiv (\sin^2 2\theta_1, \Delta m_1^2)$ and $\theta_2 \equiv (\sin^2 2 \theta_2, \Delta m^2_2)$. The network, called a direct amortized neural likelihood estimator (DNRE) \cite{10.1609/aaai.v38i18.30018}, takes as inputs $(x, \theta, \theta')$, where $\theta, \theta \overset{\mathrm{iid}}{\sim} p_\theta$ for prior $p_\theta$, and $x \sim p_{x | \theta} (\theta)$ for probabilistic simulator $p_{x | \theta} (\theta)$. A classifier network $f(x , \theta_1, \theta_2)$, trained to predict whether $x \sim p_{x | \theta} (\theta_1)$ or $x \sim p_{x | \theta} (\theta_2)$, provides a point estimate of the likelihood ratio $r(x, \theta_1, \theta_2) = p(x | \theta_1) / p (x|\theta_2)$, 

\begin{equation}
    \frac{p(x|\theta_1)}{p(x|\theta_2)} \approx \hat{r} (x, \theta_1, \theta_2) = \frac{f(x, \theta_1, \theta_2)}{1- f(x, \theta_1, \theta_2)},
\end{equation}
under regularity assumptions on the classifier network $f$.

The Neyman-Pearson lemma states that the likelihood ratio test is the most powerful method for choosing between a simple null hypothesis and a simple alternative hypothesis \cite{neyman1933problem}, and the Feldman-Cousins specification of CL construction requires an ordering criterion of the test statistic to be met; that is, the test statistic must be optimal at the best-fit $\hat{\theta}$ \cite{feldman-cousins}. To determine the best fit point used in the test statistic, we use a maximum \textit{a priori} estimate from a posterior distribution estimated with a second neural network called a sequential neural posterior estimator (SNPE-C) \cite{2019arXiv190507488G}, empirically determined to be well-poised for such a task \cite{Villarreal:2025gux}. SNPE-C is a normalizing flow which learns the posterior distribution $p(\theta | x)$ over physics parameters $\theta$ for experimental pseudo-data $x$. When used in tandem with likelihood ratio estimation, trials-based CLs constructed with a surrogate test statistic demonstrably meet expected frequentist coverage guarantees. Such a procedure can be used to build confidence levels and associated experimental sensitivities which are both probabilistically meaningful and efficient to compute \cite{10.1088/2632-2153/ae040c}.

\par The pseudo-data used for network training were generated from physics parameters distributed log-uniformly on a $50 \times 50 \times 50$ grid in $(U_{e4}, U_{\mu 4}, \Delta m_{41}^2)$, with $U_{e4}$ and $U_{\mu 4}$ spanning $[0.001, 1/\sqrt{2}]$, and $\Delta m_{41}^2$ spanning $[0.01, 100]\,\text{eV}^2$. We exclude the parameter space $U_{\mu 4}, U_{e 4} > 1/\sqrt{2}$ from analysis since it is strongly disfavored by oscillation experiments \cite{Diaz:2019fwt}, and is largely disallowed from unitarity constraints. We generate $1000$ pseudo-experiments per parameter point, and pass the physics parameters $\log_{10} U_{e4}$, $\log_{10} U_{\mu 4}$, and $\log_{10} \Delta m_{41}^2$ as inputs. We separately train SNPE-C on pseudo-data $x$ and parameters $\log U_{e4}$, $\log U_{\mu 4}$, and $\log \Delta m^2_{41}$. 

Together, the SNPE-C posterior estimator and DNRE likelihood-ratio estimator can be combined to construct a surrogate log-likelihood ratio $\log \hat r (x, \theta, \hat \theta)$. We use this ratio to build confidence levels using the Feldman-Cousins technique.

\subsection{Computing PG Tension with Simulation-Based Inference} \label{subsec:methods_sbitension}
The traditional PG tension (Eq. \ref{eq:MS_PG}) is a difference in raw $\chi^2$ values, but absolute $\chi^2$ values are inaccessible via the chosen network framework. To circumvent this limitation,
we first rewrite $\chi^2_{PG}$ in terms of $\Delta \chi^2$s: 

\begin{equation}
    \label{eq:chisqpg-diff}
    \begin{split}
    \chi^2_{PG} &= \chi^2_{glob, min} - \sum_r \chi^2_{r, min} \\
         &=\chi^2_{glob, min} - \sum_r \chi^2_{r, min} - (\chi^2_{glob, 0} - \sum_r \chi^2_{r, 0}) \\
     &= \Delta \chi^2_{glob} - \sum_r \Delta \chi^2_{r}
    \end{split}
    \end{equation}
where $\chi^2_0$ is evaluated at null ($U_{\mu 4} = U_{\mu e} = 0$). The term in parentheses in the second line of Eq.~\ref{eq:chisqpg-diff} vanishes because, under the null hypothesis, the sum of the individual contributions should be identical to the $\chi^2$ of the combined fit. Throughout this work, we use $\Delta \chi^2$ to refer to the difference in $\chi^2$ between the best fit and the null hypothesis.

Since $\Delta \chi^2$ is proportional to the log-likelihood ratio, it can be approximated directly using our network. We therefore define the SBI analogue of the PG tension:
\begin{equation}
    \label{eq:chisqpg-=sbi}
    \hat T_{PG}= -2(\hat r_{glob}(x, \theta_0, \hat \theta) - \sum_r  \hat r_{r}(x, \theta_0, \hat \theta))
\end{equation}
where the corresponding SBI approximation to $\Delta \chi^2$ is 
\begin{equation}
    \Delta \tilde \chi^2 = -2 \log \hat r(x, \theta_0, \hat \theta).
\end{equation}
Below we discuss two features of this approach. 

First, because $\hat r(x, \theta, \hat \theta)$ is approximated by the network, $\hat T_{PG}$ is likewise an approximation to $\chi^2_{PG}$. We would therefore expect some small deviations from the exact $\chi^2_{PG}$ due to network estimation error. To calibrate our methods, we compare $\hat T_{PG}$ to $\chi^2_{PG}$ using a toy model in Appendix \ref{sec:calibration_tension} (Fig. \ref{fig:toy-tension}) and find good agreement between the two methods, particularly when the tension is above $2.5 \sigma$ (roughly 99\% CL). Tensions below $2 \sigma$ do not indicate notable disagreement, and thus variations between the SBI-computed tension and exact PG-tension are expected in that regime.  We note that both methods in the toy model calculate the tension empirically, meaning there is a finite cap on the maximum tension achievable. We discuss limitations of such a frequentist procedure below. 

Second, as with any empirical frequentist procedure, including the classic Feldman-Cousins technique \cite{feldman-cousins}, the precision of the empirical $p$-values of $\hat T_{PG}$ and $\Delta \tilde{\chi}^2$ is limited by the number of trials evaluated. In this work, we use $48,000$ null realizations to construct empirical distributions of $\Delta \tilde{\chi}^2$ for the individual experiments and the joint fit (this number of trials would be computationally infeasible using traditional $\chi^2$ minimization, but is reasonable with the SBI-based procedure). With $48,000$ trials, the smallest nonzero empirical $p$-value is $p_{\rm emp}=1/48,000$, corresponding to a maximum resolvable significance of approximately $4.25\sigma$. Moreover, the statistical uncertainty becomes increasingly large in the tail, where only a small number of trials have test statistics exceeding the observed.

This limitation is particularly relevant for highly significant signals. For example, a fit to MiniBooNE data without $f_\pi$ corrections yielded a $7.1\sigma$ significance using Wilks' theorem (Tab.~\ref{tab:significances_wilks}), which corresponds to a $p$-value of order $10^{-12}$ and would require more than $10^{12}$ pseudo-experiments to determine empirically. Such a number of trials is infeasible even with the accelerated SBI procedure. For a sufficiently strong signal, a trials-based significance  will effectively saturate near the tail of the empirical null distribution. In this regime, the empirical procedure cannot reliably distinguish between signals whose true significances differ substantially; for example, a $4\sigma$ and a $6\sigma$ signal may both yield empirical significances near the maximum resolvable value.

This saturation can also affect the inferred tension. The PG tension depends on the relative values of the test statistic for the individual and combined fits. If the significance of one experiment, such as MiniBooNE, is saturated in the tail while the combined fit remains well within the resolvable region of the null distribution, the saturation can artificially limit the inferred significance of the individual experiment and distort the resulting tension.

To avoid over-interpreting this tail-limited regime, we treat SBI-computed significances as lower bounds whenever too few null trials exceed the observed test statistic. We adopt a threshold of $3.5\sigma$, corresponding to approximately $23$ of the $48,000$ null trials exceeding the observed value. This is because 23 exceedances correspond to a relative binomial uncertainty of roughly $ 1/\sqrt{23}$, or 20\%. Any significance calculation in this regime, and any tension calculation that depends on a significance in this regime, is therefore reported as a lower bound. 

\section{SBI-based Fits}\label{sec:fits}
\subsection{Single Experiment Fits}
\begin{figure}
    \centering
    \begin{subfigure}{\linewidth}
        \includegraphics[width=\linewidth]{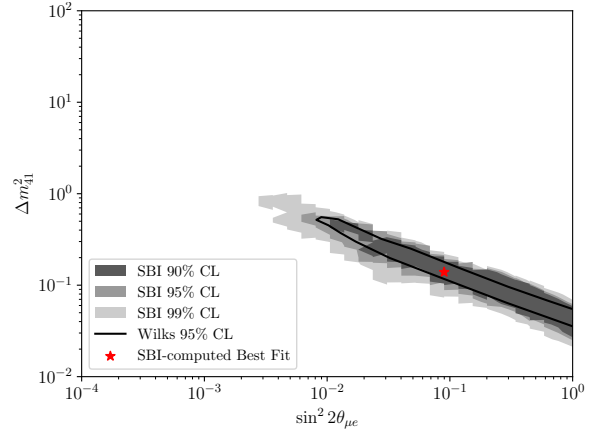}
    \caption{Fit to MiniBooNE (without $f_\pi$ corrections), with the Wilks'-based 95\% CL overlaid for comparison, since a corresponding published fit does not exist. The SBI-computed best fit is $U_{e4} = 0.41$, $U_{\mu 4} = 0.41$, and $\Delta m_{41}^2 = 0.115$ eV$^2$.}
    \label{fig:sbi-singlefitsminiboone}
    \end{subfigure}
    \hfill
    \begin{subfigure}{\linewidth}
        \includegraphics[width=\linewidth]{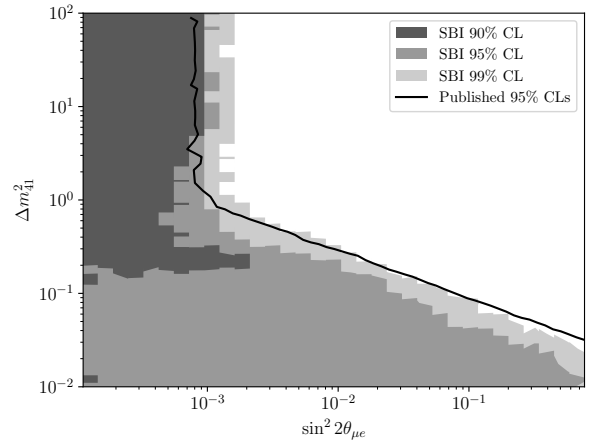}
    \caption{Fit to MicroBooNE, using data from both BNB and NuMI beamlines, with the published result using the CL$_s$ method taken from Ref. \cite{MiniBooNE:2022emn} overlaid. The SBI-computed best fit (not within plotting region) is $U_{e4}= 0.18$, $U_{\mu 4} = 0.002$, and $\Delta m_{41}^2 = 3.39$ eV$^2$.}
    \label{fig:sbi-singlefitsmicroboone}
    \end{subfigure}
    \hfill

    \caption{SBI-based fits to single experiments.}
    \label{fig:sbi-singlefits}
\end{figure}

SBI-based fits to MiniBooNE and MicroBooNE are shown in Fig. \ref{fig:sbi-singlefits}. For MiniBooNE, we compare the SBI-based CLs to a Wilks'-based CL and observe strong agreement. 

The interpretation of the SBI-based MicroBooNE fit is more nuanced. At high CLs, we observe good agreement with the published fit and the SBI-computed result. Moreover,  the SBI-computed best fit (Fig. \ref{fig:sbi-singlefitsmicroboone}, caption) is almost identical to the best-fit found in the Wilks'-based fit: $U_{e4} = 0.12$, $U_{\mu 4} = 0.001$, and $\Delta m_{41}^2 = 2.33$ eV$^2$. 

One important feature in the SBI-based fit is the emergence of a preferred region at $90\%$ CL at low $\sin^2 2\theta_{\mu e}$ and high $\Delta m_{41}^2$. The allowed parameter space in this region corresponds to high $U_{e4}$ and low $U_{\mu 4}$, and arises from the simultaneous $\nu_\mu$ excess and $\nu_e$ deficit in the BNB channel. In these experiments, the magnitude of this $\nu_\mu$ excess cannot be explained by anomalous $\nu_e \rightarrow \nu_\mu$ oscillations because the intrinsic $\nu_e$ rate is too low. Therefore, given that a high $U_{\mu 4}$ corresponds to a $\nu_\mu$ deficit, the fit finds preference for a near zero $U_{\mu 4}$. Furthermore, since the BNB channel displays a $\nu_e$ deficit, the fit finds preference for a high $U_{e4}$ and high $\Delta m_{41}^2$ to yield non-negligible $\nu_e$ disappearance (see Eq. \ref{eq:dise}). Such a preferred region can only arise in a full 3+1 fit, not an appearance-only fit, so the shape and location may appear unfamiliar.

We note that the published 3+1 MicroBooNE BNB-only fit found a similar parameter preference \cite{MicroBooNE:2022sdp}, and that a Wilks' fit to MicroBooNE BNB-only data uncovered an allowed region over the same part of parameter space in Ref. \cite{MiniBooNE:2022emn}, albeit at a lower confidence level. Thus, we do not believe the SBI-computed preferred region over this part of parameter space is surprising.

\subsection{Combined fits}\label{subsec:comb_sbi}

In Fig. \ref{fig:nofpi_comb_fit}, we present the SBI-based joint fit to MiniBooNE and MicroBooNE. In Appendix \ref{app:miniboonefpisbi}, we present the SBI-based joint fit with MiniBooNE included in the MC, consistent with the collaboration's analysis choice. From Fig. \ref{fig:nofpi_comb_fit}, as expected, the confidence levels are certainly wider than the fit to MiniBooNE-only (Fig. \ref{fig:sbi-singlefitsminiboone}), indicating how the preference for $\nu_e$ appearance is lowered once MicroBooNE data is included. 

While there is good agreement between the Wilks'-based CL and the SBI-based fit, there are notable differences between the SBI-computed CLs and the Wilks-based CLs. We attribute this to two methodological differences: first, the use of Wilks-based confidence regions rather than the fully calibrated frequentist treatment; and second, the use of SBI-estimated $\Delta\tilde\chi^2$ values in place of the exact $\chi^2$ calculation. It is very reasonable that the combination of these two differences in methodology yields the discrepancy that we observe, since Wilks' theorem tends to under-cover in the context of neutrino oscillations, as discussed in Ref. \cite{Hardin:2022qdh}, and SBI-based fits tend to be slightly more conservative when fitting to data that does not resemble simulation exactly (see Sec 4.3 of Ref. \cite{10.1088/2632-2153/ae040c}). We elaborate on these differences in Sec. \ref{sec:discussion}. 

The ``spottiness" of the 99\% CL, particularly on the left border of the $\sin^2 2\theta_{\mu e }$ space plotted, is an artifact of computing SBI-approximated $\Delta \tilde \chi^2$ values. Small fluctuations in the estimated $\Delta \tilde \chi^2$ can cause parameter points near the exclusion threshold to venture into the allowed region and therefore should not be interpreted as physical, and could likely be corrected with more trials. 
\begin{figure}[!htb]
    \centering
    \includegraphics[width=\linewidth]{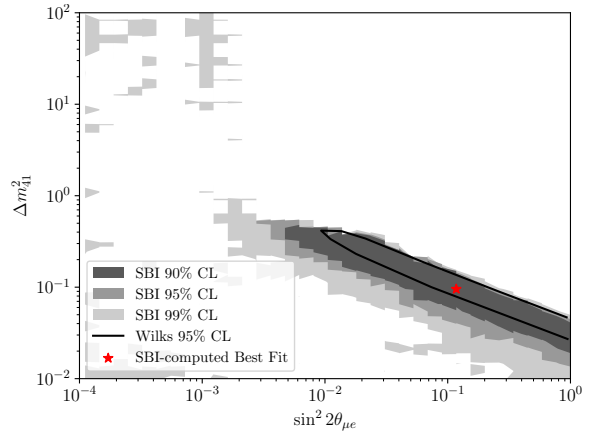}
    
    \caption{SBI-based fit to MiniBooNE and MicroBooNE, with the Wilks'-based 95\% CL overlaid in black. The $f_\pi$ factors are removed from the MiniBooNE fit. The SBI-computed best fit is $U_{e4}=0.47$, $U_{\mu 4}=0.36$, and $\Delta m_{41}^2=0.10$eV$^2$.}
    \label{fig:nofpi_comb_fit}
\end{figure}

Using the trials-based SBI procedure, the MiniBooNE-only fit finds preference for 3+1 at $\geq 3.6 \sigma$, and the combined fit finds preference for 3+1 at $2.7 \sigma$. The SBI-measured PG tension between the two experiments is $\geq 2.5 \sigma$, indicating moderate disagreement between the two experiments under the common 3+1 model. 

\section{Discussion: Comparison of the SBI- and Wilks'-based approach}\label{sec:discussion}
While we cannot explicitly quantify the disagreement between the analytic Wilks'-based method for computing tension and the SBI-approximated, trials-based method, we discuss here some differences in the methods that would result in different tensions. First, for such a strong oscillatory signal, the behavior of the test statistic may be ``non-Wilksian" \cite{Hardin:2022qdh}. Second, the finite number of trials evaluated in the SBI-based approach caps the maximum significance at $4.25 \sigma$, leading to a saturation effect described in Sec. \ref{subsec:methods_sbitension}. A third possible source of discrepancy is the behavior of the SBI neural ratio estimator when evaluating data that do not resemble realizations seen in network training. In past papers (Ref. \cite{10.1088/2632-2153/ae040c}), we have observed that if there are data/MC differences, the SBI-based method tends to penalize the test statistic more than in a Wilks'-based fit. That may also be the case here.

Because of the advantages and limitations of each approach, it is important to conduct statistical tests using both an analytic Wilks'-based approach and a trials-based SBI approach. The two statistical frameworks are complementary; SBI can supplement the coverage issues associated with Wilks' theorem, while Wilks' theorem can analytically assess signal significance when too many trials are required of SBI. Although the two methods may yield different statistics, the qualitative interpretation of the significance and tension is consistent between the two methods. In this case, both methods find a statistically significant tension between MiniBooNE and MicroBooNE.
\section{Conclusion}\label{sec:conclusion}

In this paper, we have established a unified treatment of MiniBooNE and MicroBooNE data, using consistent fitting methods and a consistent treatment of data-driven corrections. This work presents the first combined MiniBooNE–MicroBooNE fit to the 3+1 sterile-neutrino model using both MicroBooNE BNB and NuMI datasets and without the inclusion of $f_\pi$ in the MiniBooNE fit. Using the trials-based SBI procedure, a joint fit to both datasets finds a preference to 3+1 of $2.7 \sigma$, and the two experiments exhibit a $\geq 2.5 \sigma$ PG tension, indicating that their preferred regions of parameter space are not fully compatible within the 3+1 model. This is particularly notable because both experiments operate along the same BNB neutrino beamline and are therefore expected to share many beam-related and environmental systematic uncertainties.

In this work we have also presented a novel method for evaluating tension between datasets using SBI under a trials-based approach. This method provides a computationally efficient way to quantify the compatibility of the two datasets without relying on asymptotic assumptions. The methods we have demonstrated in this analysis can be directly applied to a full 3+1 global fit to assess the tension between datasets sensitive to neutrino appearance and disappearance, as well as other scenarios in particle physics where multiple experiments exhibit tension with a common theoretical model. Coupled with Ref. \cite{10.1088/2632-2153/ae040c} and Ref. \cite{Villarreal:2025gux}, this work establishes a complete framework for conducting an entirely SBI-based global fit, incorporating frequentist and Bayesian inference as well as methods for quantifying tension between datasets. This is particularly important because SBI-based techniques enable global fits involving more complex sterile neutrino scenarios, which have gained increasing interest given that, as shown here, a minimal 3+1 framework cannot simultaneously describe the data from multiple experiments.

We make the following recommendations regarding data releases that will allow the points examined here to be further solidified:
\begin{itemize}
\item A release of MicroBooNE event-by-event Monte Carlo data with the true $L$, true $E$ and reconstructed $E$ for each event for the simulation they use for analysis.    This has been the standard for oscillation data releases in the past, including all MiniBooNE releases,  and is required for the most precise global fits.   
\item A release of MicroBooNE parent information for neutrinos in the event-by-event Monte Carlo data.  In our data release, we supply this information for MiniBooNE, provided by and released with permission by the collaboration. 

\end{itemize}

\bmhead{Acknowledgements}
We are grateful to Mike Shaevitz at Columbia University and Matheus Hostert of University of Iowa for helpful discussions throughout the paper.  JW, JV, JH, and JC thank MIT for support on this project. AS thanks Texas A\&M University for support on this project. This material is based upon work supported in part by the National Science Foundation Graduate Research Fellowship under Grant No. 2141064. We thank MiniBooNE for providing internal MC files upon request. 

\appendix

\section{Impact of $f_\pi$ on BNB-only fits }\label{sec:BNBfpi}

In this appendix, we consider all combinations of adding and removing $f_\pi$ from MiniBooNE \textit{and} MicroBooNE BNB data to understand the consequences in the fits and tension. Since no parent information is available for MicroBooNE, we cannot perform the same event-by-event scaling that was used in the MiniBooNE BNB analysis. Instead, we apply the same factors as an approximate correction, assuming that the fraction of events coming from $\pi^+$ is the same function of reconstructed energy in MiniBooNE and MicroBooNE, and that this fraction is relatively constant per bin. 

\begin{figure*}[htb]
    \centering
    \begin{subfigure}{0.48\textwidth}
        \includegraphics[width=\linewidth]{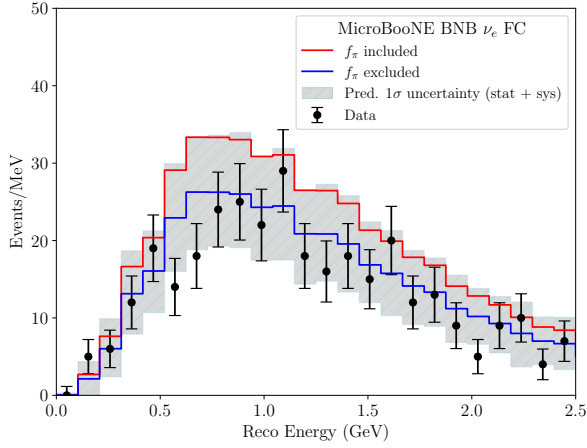}
    \end{subfigure}
    \hfill
    \begin{subfigure}{0.48\textwidth}
        \includegraphics[width=\linewidth]{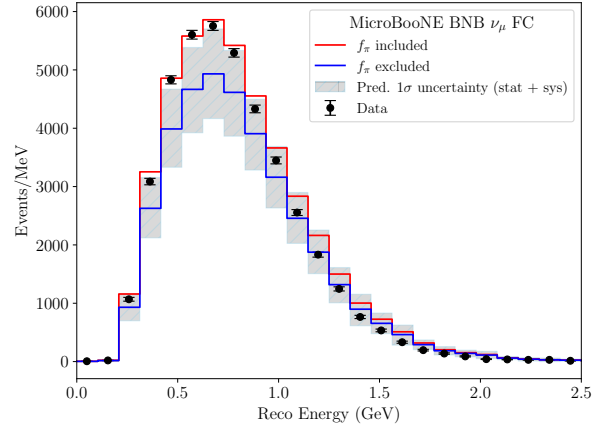}
    \end{subfigure}
    \hfill
    \caption{Null MicroBooNE expectation for selected $\nu_e$ and $\nu_\mu$ channels with and without $f_\pi$ included in the fit. Predicted $1 \sigma$ combined statistical and systematic uncertainty bands for the prediction without $f_\pi$ corrections (see Fig. \ref{fig:BNBdata_dists}) are shown in gray.}
    \label{fig:fpi_microboone}
\end{figure*}

Reconstructed energy distributions for MicroBooNE with and without $f_\pi$ are shown in Fig. \ref{fig:fpi_microboone}. The inclusion of $f_\pi$ correction improves the agreement in the $\nu_\mu$ channel, particularly at lower energies, where an excess is no longer present. However, the $\nu_e$ deficit increases when $f_\pi$ is included, significantly worsening the fit to null.

Fit statistics for each combination of experiment with and without $f_\pi$ are shown in Table \ref{tab:fitstatistics}. We first note that the $\chi^2/\mathrm{dof}$ for MiniBooNE with $f_\pi$ is unusually low. Such a low $\chi^2$ implies that the data matches expectation much better than pseudo-experiments, unsurprising given $f_\pi$ was added to match $\nu_\mu$ data with expectation. Without $f_\pi$ included, the MiniBooNE fit finds a $\chi^2/\mathrm{dof}$ of $\sim 1$, indicating a good fit, and lending credence to the decision that $f_\pi$ should be omitted from MiniBooNE in a full 3+1 fit (see Sec \ref{sec:fpiconsistent}). We also note that the $\chi^2/\mathrm{dof}$ for MicroBooNE with or without $f_\pi$ corrections is similarly low, as discussed in Sec \ref{sec:microboone}. 

While including $f_\pi$ from MiniBooNE drastically lowers the signal significance, including $f_\pi$ in the MicroBooNE BNB fit slightly raises the $\Delta \chi^2$ due to increased disagreement with null in the $\nu_e$ channel, yet MicroBooNE remains consistent with null. In fact, the exclusions drawn from a Wilks' fit to MicroBooNE BNB+NuMI data have minor, if any, changes when $f_\pi$ is included in the BNB prediction. 

Table \ref{tab:fitstatistics} shows that if $f_\pi$ is included in a consistent way in a joint fit to both experiments' BNB analyses, the appearance anomaly remains at $3.9 \sigma$. If $f_\pi$ is omitted from both experiments, the appearance anomaly increases to $5.9 \sigma$. The $>5 \sigma$ strength of the 3+1 preference from a joint fit to both BNB datasets further motivates the need to understand the source of the $\nu_\mu$ disagreement.

\begin{table*}
    \centering
    \caption{Fit statistics using a traditional $\chi^2$ for MiniBooNE and MicroBooNE (BNB-only) in a full 3+1 fit. Note: the label ``($f_\pi$)" denotes the inclusion of $f_\pi$ factors.}
        
    \begin{tabular*}{\linewidth}{@{\extracolsep{\fill}} lcc}
        \toprule
        \textbf{Dataset(s)} & $\chi^2_{min}/N$ dof & $\Delta \chi^2$ (wrt null)/3 dof \\
        \midrule
        MiniBooNE ($f_\pi$)  &  21.5/35 & 29.7  \\
        MiniBooNE & 34.1/35 &  58.6\\
        MicroBooNE ($f_\pi$)  &  52.61/97 &  3.0 \\
        MicroBooNE &  64.98/97 &   2.4 \\
        MiniBooNE ($f_\pi$) +MicroBooNE ($f_\pi$) &  85.97/135 & 21.0 \\ 
        MiniBooNE ($f_\pi$) +MicroBooNE &  99.16/135 & 19.51 \\ 
        MiniBooNE+MicroBooNE ($f_\pi$) &  106.78/135 & 43.32\\ 
        MiniBooNE+MicroBooNE  &  120.68/135 & 41.94 \\ 
        \bottomrule
    \end{tabular*}
    \label{tab:fitstatistics}

\end{table*}

We report the tension between MiniBooNE and MicroBooNE (BNB-only) for all combinations of experiments with and without $f_\pi$ in Table \ref{tab:fpitension}, assuming the PG tension follows a $\chi^2_{3 \mathrm{dof}}$ distribution. The tension is highest when $f_\pi$ is omitted from both MiniBooNE and MicroBooNE fits, where a combined fit cannot find a sufficient set of parameters that describe the increased MiniBooNE $\nu_e$ excess and the increased MicroBooNE $\nu_e$ deficit.   However, the tension is always $>2\sigma$ regardless of $f_\pi$ inclusion.  Thus, the differing $f_\pi$ treatments is not the only cause of internal disagreement between MiniBooNE and MicroBooNE.
\begin{table}[ht]
\centering
\caption{Wilks'-based tension between MiniBooNE and MicroBooNE BNB-only fits, following the ``$(f_\pi)$'' notation in Table \ref{tab:fitstatistics}.}
\label{tab:fpitension}
\begin{tabular}{lc}
\toprule
\textbf{Dataset(s)} & \textbf{Tension  ($\sigma$)} \\
\midrule
MiniBooNE ($f_\pi$) v. MicroBooNE ($f_\pi$) & 2.63 \\
MiniBooNE ($f_\pi$) v. MicroBooNE &  2.78 \\
MiniBooNE v. MicroBooNE ($f_\pi$) &  3.55 \\
MiniBooNE v. MicroBooNE &  3.65 \\
\bottomrule
\end{tabular}

\end{table}

\section{Wilks-based 3D Fits} \label{app:wilks3D}
\begin{figure*}[htb]
    \centering
        \includegraphics[width=0.95\linewidth]{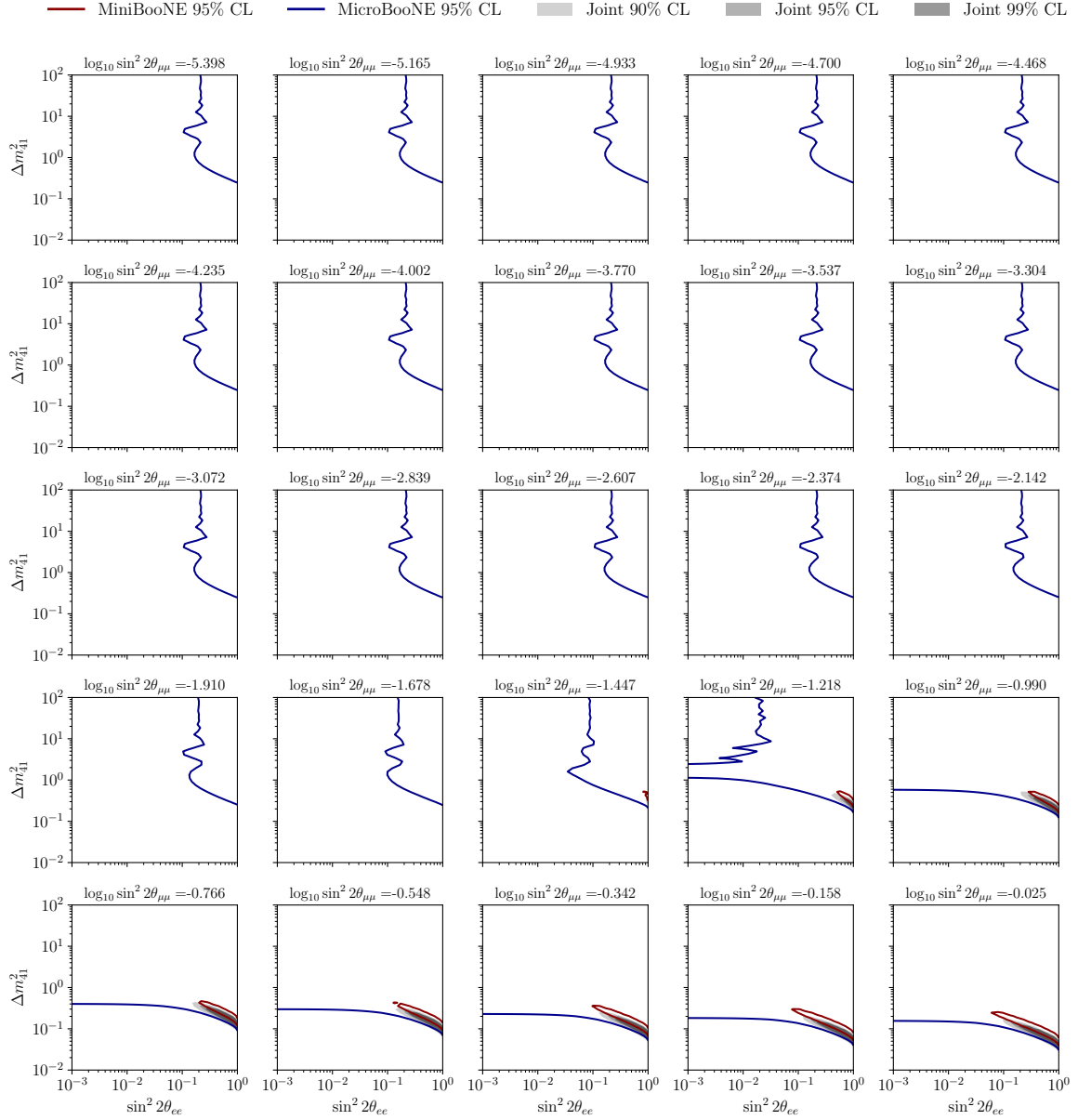}
    \caption{$\nu_e$ disappearance plots for the joint fit, at various fixed values of $\sin^2 2\theta_{\mu \mu}$, with MiniBooNE (without $f_\pi$) and MicroBooNE-only overlaid.}
    \label{fig:Wilks_ps_ee}
\end{figure*}

\begin{figure*}[htb]
    \centering
    \includegraphics[width=0.95\linewidth]{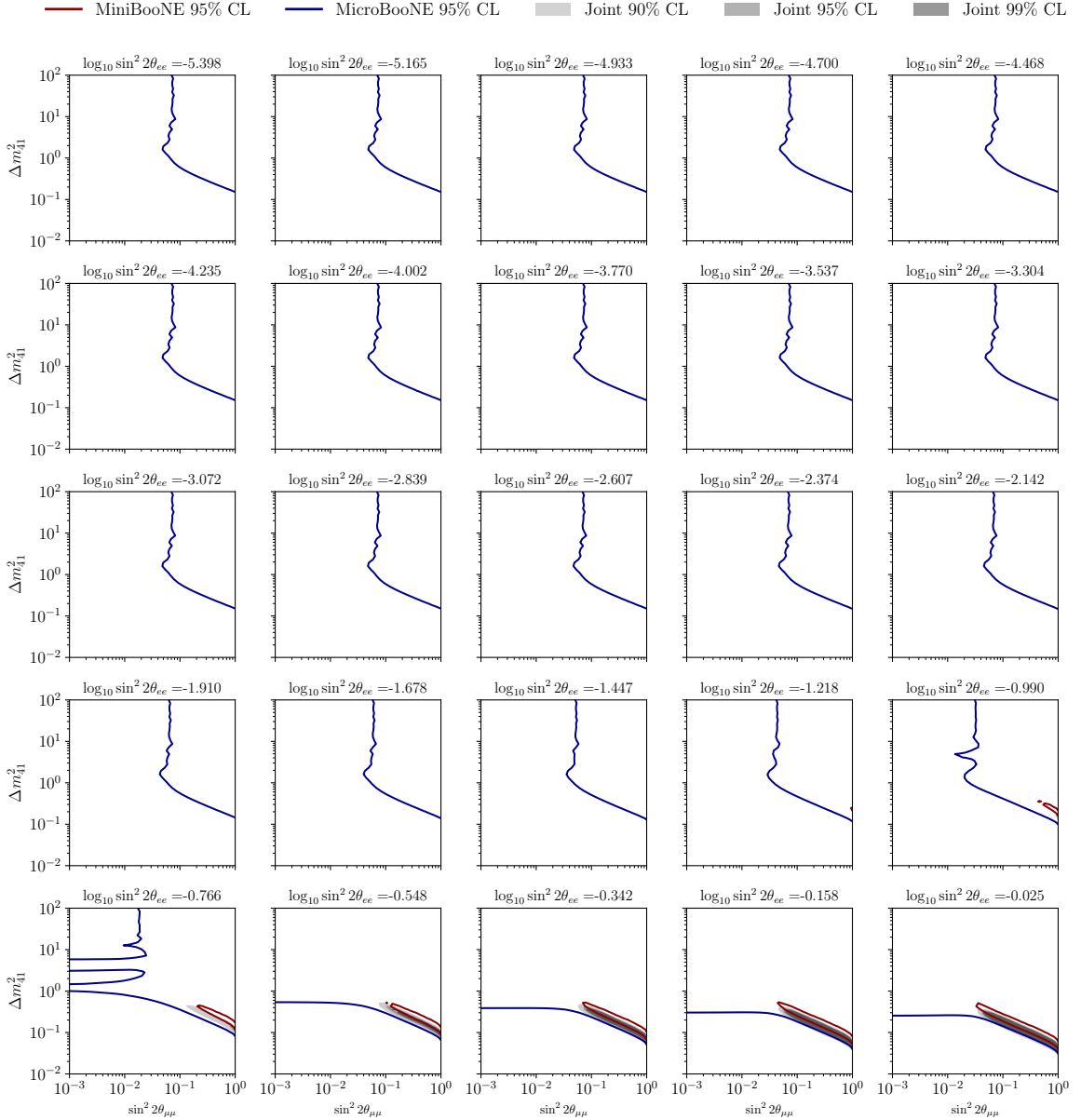}
    \caption{$\nu_\mu$ disappearance plots for the joint fit, at various fixed values of $\sin^2 2\theta_{ee}$, with MiniBooNE (without $f_\pi$) and MicroBooNE-only overlaid.}
    \label{fig:Wilks_ps_uu}
\end{figure*}

Cross sections of the three-dimensional 3+1 parameter space from a joint fit to MiniBooNE and MicroBooNE data assuming Wilks' theorem, shown in Figs. \ref{fig:Wilks_ps_ee} and \ref{fig:Wilks_ps_uu}. 


\section{Calibration checks using a toy model}\label{sec:calibration}
In this section, we consider a toy experiment sensitive to $\nu_e$ appearance, using exactly the same toy model as that considered in the Feldman and Cousins paper (Ref. \cite{feldman-cousins}) as well as Ref. \cite{10.1088/2632-2153/ae040c}. The experiment is modeled after accelerator-based experiments, where pion-decays produce muon neutrinos at baselines 600-1000m from the detector, with energy range 10-60 GeV. We fit to two parameters, $\sin^2 2 \theta_{\mu e}$, and $\Delta m_{41}^2$, over a $30 \times 30$ grid spanning $\sin^2 2 \theta_{\mu e} \in [10^{-4}, 0]$ and $\Delta m_{41}^2 \in [1, 10^3]$ eV$^2$ uniformly in log space. 

Our exact $\Delta \chi^2$ for comparison to the SBI-approximated log-likelihood ratio takes the form 
\begin{equation}
    \Delta \chi^2 = 2 \sum_i [\mu_i - \mu_{best} + n_i \ln \frac{\mu_{best} + b_i}{\mu_i + b_i}]
\end{equation}
for expectation $\mu$, data $n$, and background $b$. 

To fit using SBI, we train our neural classifier on the two parameters of interest and train a simple posterior estimator built directly from the classifier, as was done in Ref. \cite{10.1088/2632-2153/ae040c}, to find the maximum \textit{a posteriori}. 

Code to reproduce our results is given in Ref. \cite{toygithub}.
\subsection{Calibration of SBI-based fitting}
\begin{figure}
    \centering
    \begin{subfigure}{\linewidth}
    \includegraphics[width=0.95\linewidth]{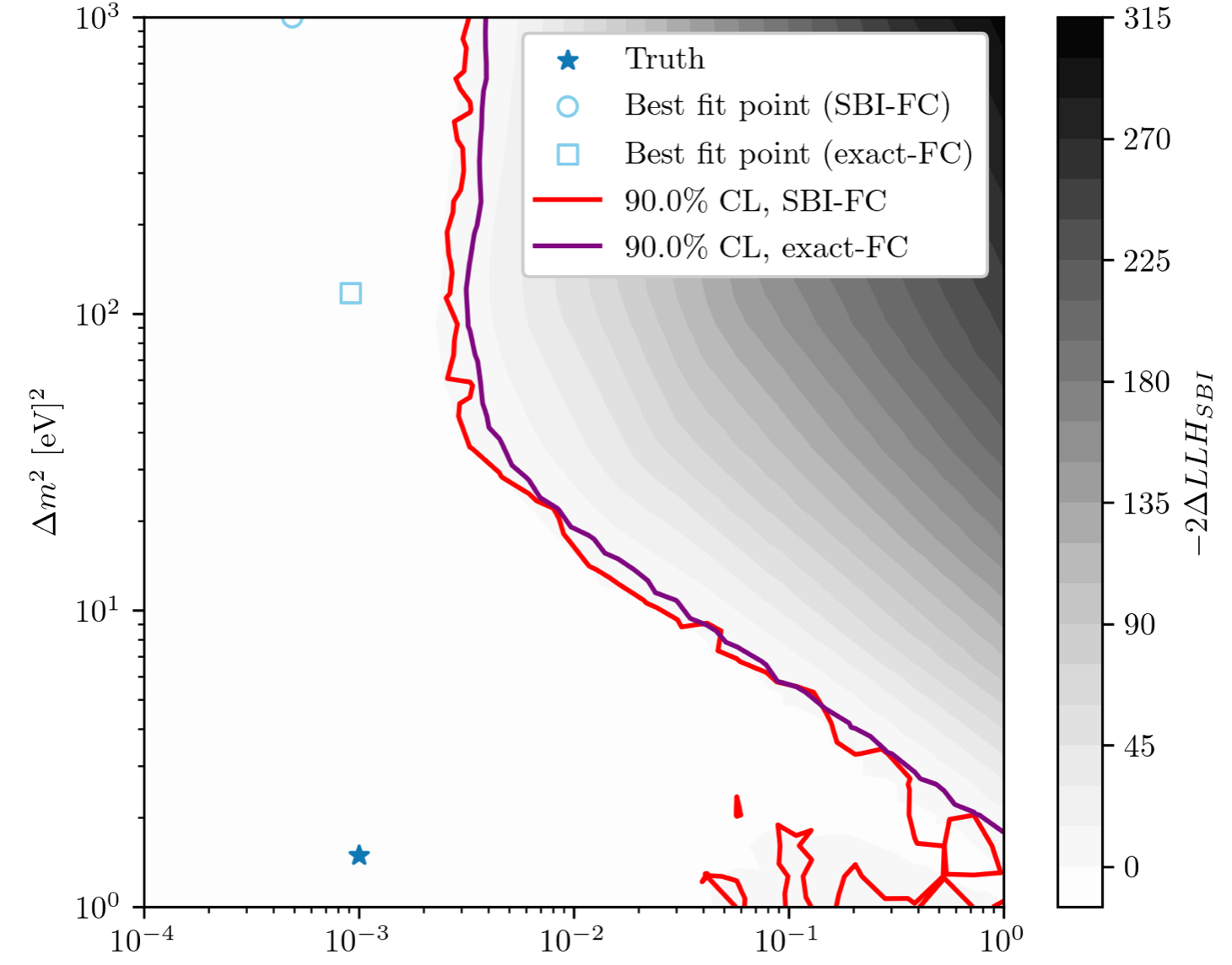}
    \end{subfigure}
    \hfill
    \begin{subfigure}{\linewidth}
    \includegraphics[width=0.95\linewidth]{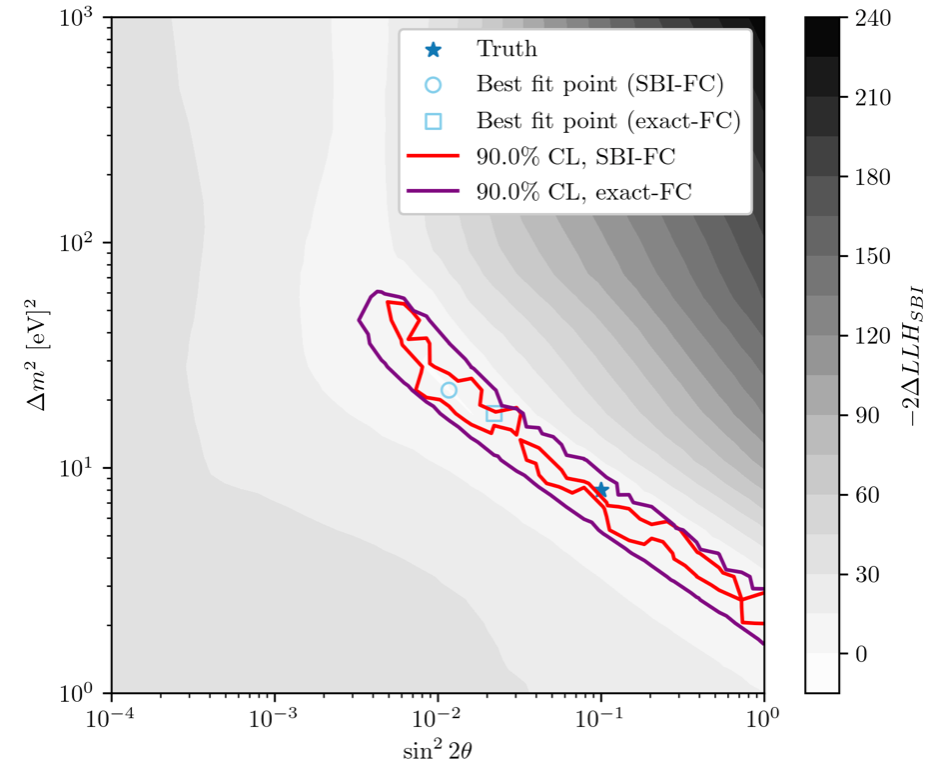}
    \end{subfigure}
    \hfill
    \caption{SBI-based fits to a pseudo-experiment generated from null-space (top) and signal-space (bottom) for a toy $\nu_e$ appearance experiment. Comparisons to a Feldman-Cousins-computed CL using an exact $\chi^2$ are also shown.}
    \label{fig:fits-toy-realizations}
\end{figure}

The calibration of SBI-based fitting was performed in Ref. \cite{10.1088/2632-2153/ae040c}, for which a toy model was built. Here we present and expand upon the results of the toy model.
In Fig. \ref{fig:fits-toy-realizations}, we compare the confidence levels computed using Feldman Cousins with an exact $\chi^2$ to that with an SBI-approximated $\Delta \chi^2$. We present fits to both a null-like and a signal-like pseudo-experiment, demonstrating their agreement for both excluded and allowed regions of the parameter space. 

\begin{figure}
    \centering
    \includegraphics[width=\linewidth,trim=0.3cm 0cm 0.2cm 0.4cm,clip]{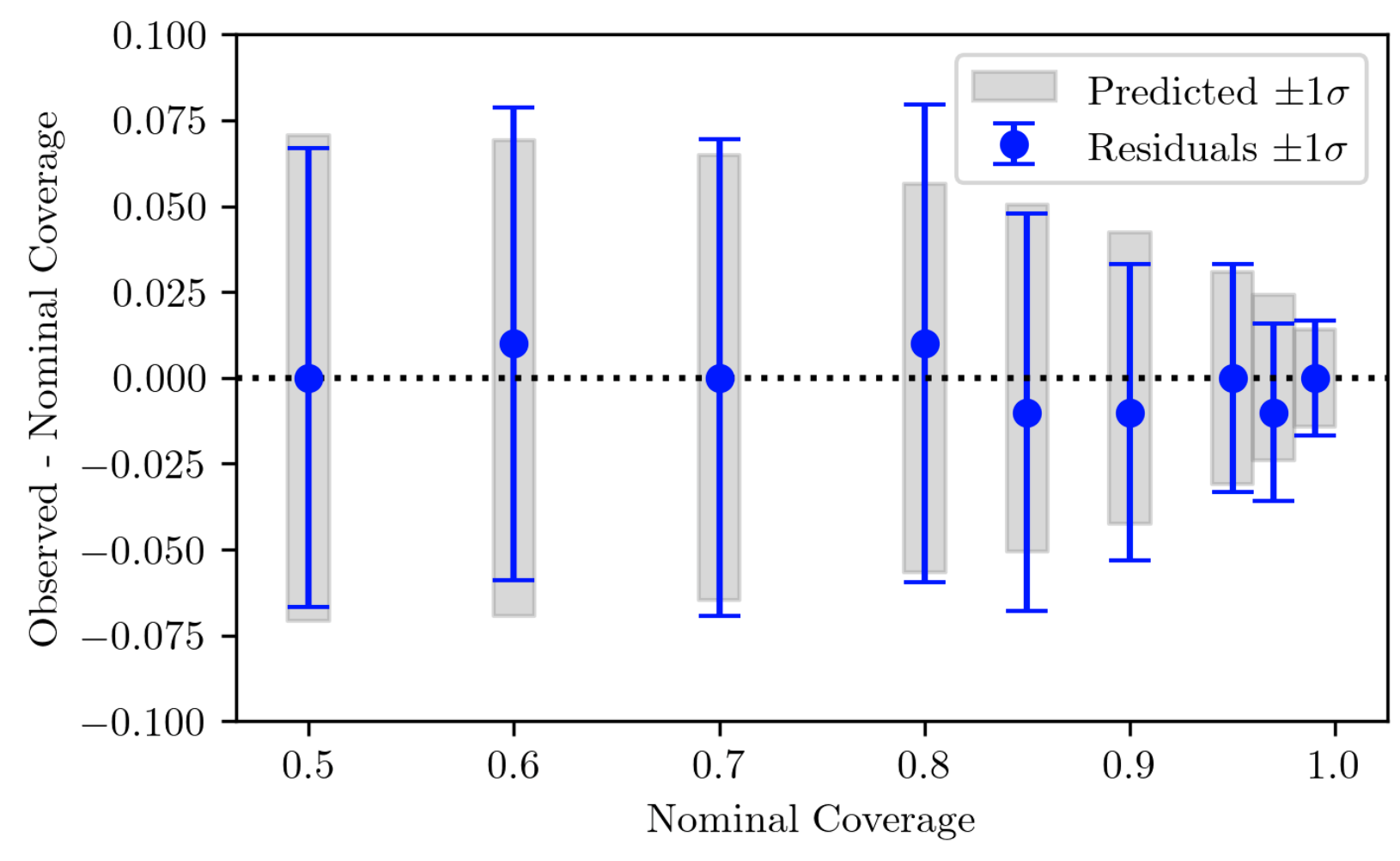}
    \caption{Difference between observed and nominal coverage for 50 SBI-evaluated pseudo-experiments generated at 25 randomly sampled parameter space gridpoints. The predicted residual distribution has a binomial variance, i.e. $\hat r \sim \mathcal{N}(0, \alpha(1-\alpha)/M)$, where $M$ is the number of trials.}
    \label{fig:toy-coverage}
\end{figure}

We also demonstrate appropriate coverage of the SBI-computed confidence levels in Fig. \ref{fig:toy-coverage}. 

\subsection{Calibration of SBI-computed tension}\label{sec:calibration_tension}
In this section, we use our toy model to demonstrate the calibration of SBI-computed $\hat T_{PG}$ using the traditional PG tension $\chi^2_{PG}$. For this toy model, the computational burden of evaluating a large number of trials is alleviated, and we can compare the tensions measured by the two methods empirically. 

\begin{figure}
    \centering
    \includegraphics[width=\linewidth,trim=0.3cm 0cm 1.3cm 1.2cm,clip]{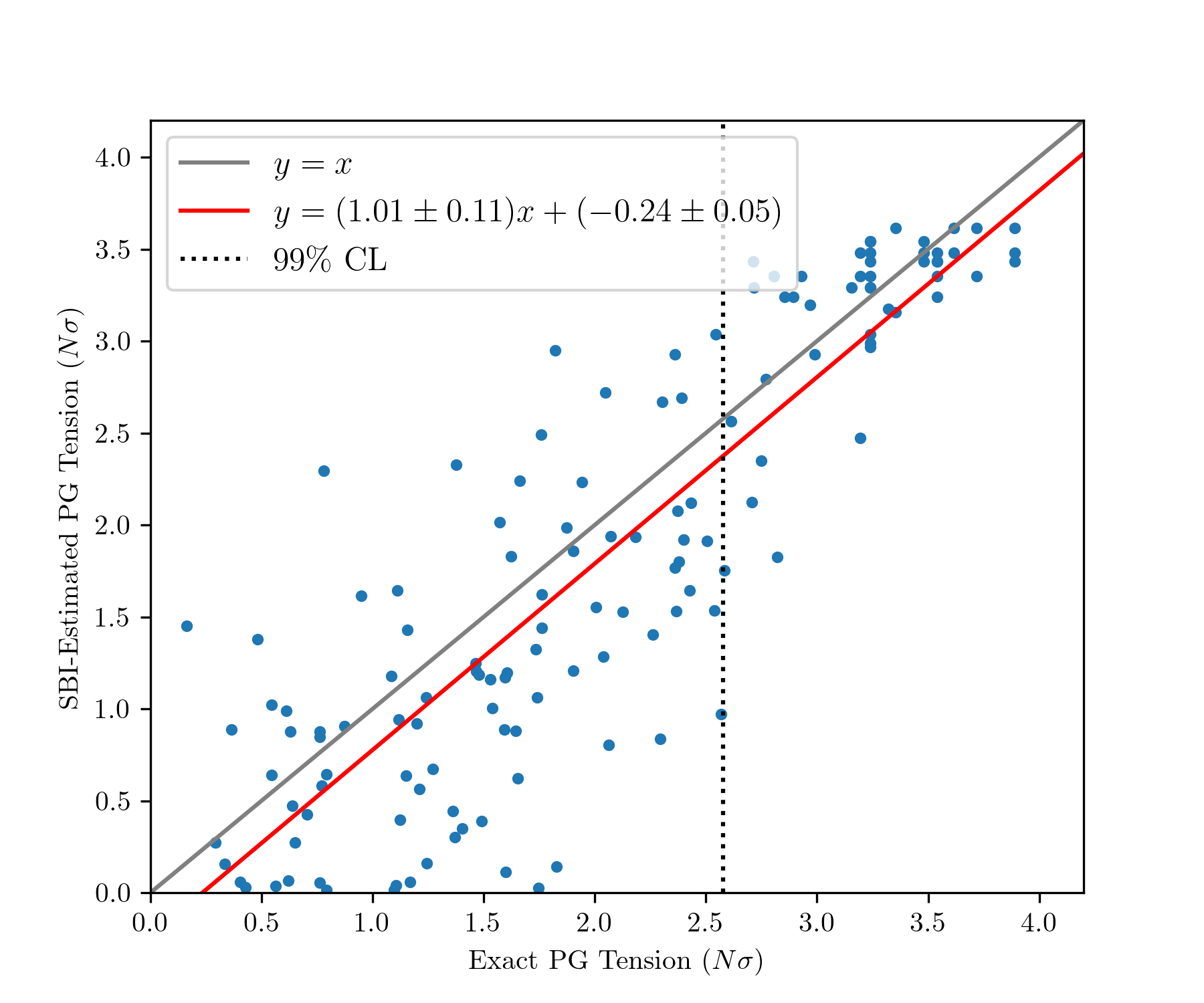}
    \caption{Empirically-calculated tension significances for pseudo-experiments generated with experiment A under the null hypothesis and experiment B in the signal-like parameter space. We compare the empirical significance of SBI-evaluated $\hat{T}_{PG}$'s with the empirical significance of exact $\chi^2_{PG}$'s. The 99\% CL on the exact PG tension is shown as the dotted line.}
    \label{fig:toy-tension}
\end{figure}

We consider two experiments (A and B) with identical statistical and systematic uncertainty, and simulate ``tense" realizations from signal-like space for experiment A and from null-like space for experiment B. We compare the two methods by computing the empirical $p$-value of each realization with respect to the corresponding null distribution, which is comprised of 10,000 trials. The results are shown in Fig. \ref{fig:toy-tension}. A linear fit to data is shown in red, with errors on each fit parameter shown in the legend. While there seems to be a slight systematic disagreement between the SBI-computed tension and the exact PG tension, indicated by the deviations of the fit line from $y=x$, the uncertainty on the fitted parameters encompass $y=x$, and we conclude there is good agreement between the two methods. 

Moreover, it is important to note that we do not expect the two methods to agree exactly. For one, the SBI-estimated tension is constructed from approximated $\Delta \chi^2$s, each of which carries their own network error. Second, although $\hat T_{PG}$ and $\chi^2_{PG}$ quantify the same underlying notion of parameter-space tension, they are obtained through different statistical constructions. The former uses SBI-based estimates of the likelihood ratios, while the latter is calculated directly from the exact likelihood. Consequently, small differences between the two methods are expected and do not necessarily indicate a failure of the SBI approach. Nevertheless, for empirical tensions calculated from the exact $\chi^2_{PG}$ that are above the 99\% CL, we note good agreement with the empirical SBI-based tension. 

We note here that all statistics presented in Fig. \ref{fig:toy-tension} are computed empirically, and thus both suffer from the same saturation effect. Therefore the saturation behavior does not appear in a comparison between the two methods and is not evident in Fig. \ref{fig:toy-tension}.

\section{SBI-based fits Adopting the Same $f_\pi$ Analysis Choices as Each Collaboration} \label{app:miniboonefpisbi}
\begin{figure}[!htb]
    \centering
    \includegraphics[width=\linewidth,trim=0.3cm 0cm 1.3cm 1.2cm,clip]{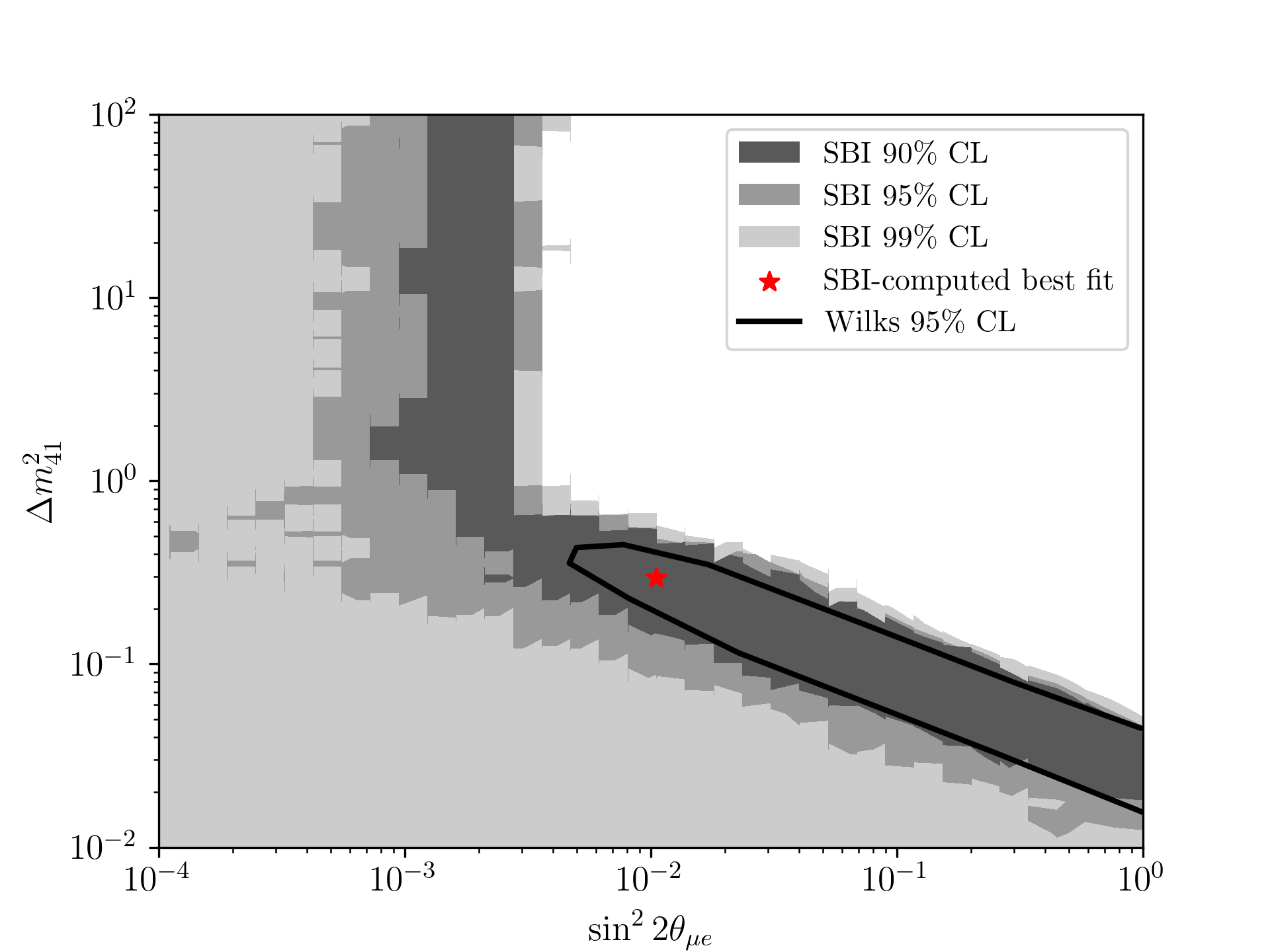}
    
    \caption{SBI-based fit to MiniBooNE and MicroBooNE, with the Wilks'-based 95\% CL overlaid in black. The SBI-computed best fit is $U_{e4} = 0.28, U_{\mu 4} = 0.19$, and $\Delta m_{41}^2 = 0.3$eV$^2$.}
    \label{fig:joint_fit_sbi_fpiminiboone}
\end{figure}

A joint fit to MiniBooNE with the factors of $f_\pi$ included and MicroBooNE is shown in Fig. \ref{fig:joint_fit_sbi_fpiminiboone}. At low values of $\Delta m_{41}^2$, there is strong agreement between the SBI-based fit and the Wilks’-based fit. The SBI-based allowed regions at 90\% and 95\% CL can be understood as a normalization effect, in which rapid oscillations (corresponding to high $\Delta m_{41}^2$ ) effectively resemble an overall normalization set by a particular value of $\sin^2 2 \theta_{\mu e}$. The fit prefers 3+1 at $1.8\sigma$. Note that the joint fit without $f_\pi$ corrections for MiniBooNE finds a greater preference for 3+1 due to an increased MiniBooNE low energy excess. The SBI-computed tension between MiniBooNE with $f_\pi$ corrections and MicroBooNE is $\geq 3.2 \sigma$. 

\clearpage
\bibliography{bibliography}% 

\end{document}